\documentclass[showpacs,aps,pra,superscriptaddress,longbibliography]{revtex4-2}
\usepackage{geometry}
\usepackage{amsfonts}
\usepackage{graphicx}
\usepackage{breakcites}
\usepackage{amsmath}
\usepackage{hyperref}
\usepackage{amssymb}
\usepackage{cleveref}
\usepackage{mathtools}
\usepackage{physics}
\usepackage{color}
\usepackage{ulem}
\usepackage[utf8]{inputenc}
\usepackage{comment}
\begin{document}
\title{Path-integral approach to the thermodynamics of bosons with memory: Partition function and specific heat}

\author{T. Ichmoukhamedov}
\author{J. Tempere}
 
\affiliation{TQC, Departement Fysica, Universiteit Antwerpen, Universiteitsplein 1, 2610 Antwerpen, Belgium}
\date{\today}

\begin{abstract}
For a system of bosons that interact through a class of general memory kernels, a recurrence relation for the partition function is derived within the path-integral formalism.
This approach provides a generalization to previously known treatments in the literature of harmonically coupled systems of identical particles.
As an example the result is applied to the specific heat of a simplified model of an open quantum system of bosons, harmonically coupled to a reservoir of distinguishable fictitious masses.  
\end{abstract}


\maketitle

\section{INTRODUCTION}

One of the peculiar features of the path integral approach in statistical mechanics is the appearance of retarded interactions between particles, after integrating out the degrees of freedom of the environment that the particles couple to.
This most famously appears in Feynman's variational treatment of the polaron problem \cite{Feynman1955}, where the effective action functional of an electron interacting with a bath of phonons is obtained by integrating out the phonon degrees of freedom. 
The resulting action functional describes the electron interacting with itself at previous times through an interaction term that is moreover non-quadratic in the electron coordinate, prohibiting an analytical solution.
A variational solution can be found by proposing a model action functional where this interaction term is replaced by a quadratic retarded interaction of the electron with itself, mediated by some memory kernel that depends on variational parameters. 
In Feynman's initial treatment of the problem, the model action is obtained by integrating out a harmonically coupled fictitious particle to the electron \cite{feynman1998statistical}, which yields a memory kernel as a function of two variational parameters. 

This approach has since then known various extensions towards either generalizations of the model action, applications to multiple particles or entirely different physical systems \cite{Feynman1963,Caldeira1983}.
In the context of a single solid state polaron, instead of obtaining a model action by integrating out fictitious particles which inevitably restricts the variational freedom of the memory kernel, an approach directly proposing the most general quadratic single-particle model action with memory has been studied in \cite{Rosenfelder2001}. 
On the other hand extensions towards multiple particles \cite{Verbist1992,Casteels2013} or towards an arbitrary number of identical particles \cite{Klimin2004}, have so far relied on using restricted model actions that are constructed by integrating out harmonically coupled fictitious particles. 
Even in the absence of a fictitious system the canonical treatment of identical particles in the path integral formalism significantly complicates the expressions for the partition and correlation functions \cite{Brosens1997a,Brosens1997b}. 
This naturally raises the question as to how these approaches could be extended to many identical particles using action functionals with general memory kernels, yielding an all encompassing treatment. In this work we answer the first part of this question by presenting a derivation of the partition function for such a general action functional. When necessary to emphasize that the memory kernels arise from the influence of an external system we will also refer to this quantity as the reduced partition function. 

It is important to emphasize that in the context of variational models, the environment often plays merely the role of an intermediary used to obtain a variationally suitable expression for the retarded interactions. 
However, the thermodynamics that follows from the reduced partition function, interpreted as a model for an open quantum system, has been shown to display particularly interesting behavior in itself \cite{Ingold2009,Ingold2014}. As an example of an application we will show how our expression for the partition function with memory could be used to generalize the study of the specific heat for identical particles in this direction as well. 

In this work we consider the following action functional (we will work in units of $\hbar=1$):
\begin{align}
S^{(N)}[\overline{\mathbf{r}}, x , y , \overline{\boldsymbol{\kappa}}]= &  \frac{m}{2} \sum_{i}^N \int_{0}^{ \beta}  \dot{\mathbf{r}}_i(\tau)^2 d\tau +  \frac{m}{2} \sum_{i}^N \int_0^{ \beta} d\tau \int_0^{\beta} d\sigma x(\tau - \sigma)  \mathbf{r}_i(\tau) \cdot \mathbf{r}_i(\sigma) \nonumber \\
 + \frac{m}{2N} & \sum_{i, j}^N \int_0^{ \beta} d\tau \int_0^{\beta} d\sigma \left[y(\tau-\sigma)-x(\tau-\sigma) \right]  \mathbf{r}_i(\tau) \cdot \mathbf{r}_j(\sigma) -m \sum_i^{N} \int_0^{\beta} d \tau \mathbf{r}_i (\tau) \cdot \boldsymbol{\kappa}_i(\tau),
\label{action1}
\end{align}
This action contains the most general quadratic many-particle potential terms.
For the functional arguments the notations $\overline{\mathbf{r}}={\mathbf{r}_1,...,\mathbf{r}_N}$ and $\overline{\boldsymbol{\kappa}}={\boldsymbol{\kappa}_1,...,\boldsymbol{\kappa}_N}$ are used. 
This (Euclidean) action functional describes $N$ particles with mass $m$ at temperature $(k_B \beta)^{-1}$, that interact through memory kernels $x(\tau-\sigma)$ and $y(\tau-\sigma)$. 
The memory kernels generally represent the effect of some external system or medium that induces retarded interactions, and would arise after integrating out the external system coupled to the particles. However, here they are taken to be completely general and can also be defined to include harmonic trapping potentials.  
In addition we introduce a set of completely general vector source functions $\boldsymbol{\kappa}_i(\tau)$, which may represent time-dependent external forces on the particles, but will mainly prove to be useful for calculating expectation values. 
Expression (\ref{action1}) can be rewritten to note that each particle interacts with itself through the memory kernel $\frac{1}{N} \left( (N-1)x(\tau-\sigma) + y(\tau-\sigma) \right)$ and with any other particle through the memory kernel $\frac{1}{N} \left( y(\tau-\sigma)- x(\tau-\sigma) \right)$, and hence the two can be tuned independently. 

We restrict the memory kernels to be symmetric $\left(x(\tau),y(\tau) \right)=\left(x(-\tau),y(-\tau) \right)$ and $\beta$-periodic $\left(x(\beta-\tau),y(\beta-\tau) \right)=\left(x(\tau),y(\tau) \right)$.  
These are general properties of bosonic Green's functions \cite{mahan} which are also assumed in the treatment for the single-polaron in \cite{Rosenfelder2001}, and naturally arise in systems with a harmonic coupling to an external system \cite{feynman1998statistical,Verbist1992,  Klimin2004, Tempere2009, Casteels2013, ExtendedFrohlichFeynman2019, houtput2020beyondfrohlich}. 
In addition we will assume that $\int_0^\beta x(\tau) d\tau \neq 0$ and $\int_0^\beta y(\tau) d\tau \neq 0$, so that we do not need to introduce a finite volume in our treatment - a technical step that occurs when taking the free particle limit as the harmonic oscillator frequency tends to zero.
We specifically consider three dimensional systems and in further notation $d=3$, unless specified otherwise.

The goal of this work is to obtain a recurrence relation for the partition function of bosons described by the general action functional (\ref{action1}). To provide an example we will apply our result to study the specific heat of the identical oscillator extension of the system in \cite{Ingold2009}. 
Our approach generalizes the previously known results for a system of harmonically coupled identical oscillators in \cite{Brosens1997a, Brosens1997b}, which corresponds to a specific choice of memory kernels in (\ref{action1}). 
First, in section \ref{Section 2} we will extend the calculation performed in \cite{Adamowski} to a many-particle system to obtain the distinguishable particle propagator corresponding to Eq.~(\ref{action1}). 
Next, in section (\ref{Section 3}) we will discuss which steps of \cite{Brosens1997a, Brosens1997b} need to be generalized to take memory effects for identical particles into account.
Therefore, in a way this work can be seen as an application of the methods in \cite{Adamowski} to generalize the approach in \cite{Brosens1997a}. 
Finally, in section (\ref{Section 4}) we will apply the results to consider the specific heat of an open quantum system of bosons, where the effects of the environment are represented by a harmonic coupling to fictitious masses. 

\section{Propagator}\label{Section 2}
  
Before taking the permutation symmetries of identical particles into account, first the many particle propagator for $N$ distinguishable particles has to be calculated:
\begin{equation}
K_N[x, y, \overline{\boldsymbol{\kappa}} ] \left( \overline{\mathbf{r}}_T  , \beta |  \overline{\mathbf{r}}_0, 0 \right) = \int_{\overline{\mathbf{r}}_0,0}^{\overline{\mathbf{r}}_T, \beta } \mathcal{D} \overline{\mathbf{r}} ~ e^{-S^{(N)}[\overline{\mathbf{r}}, x , y , \overline{\boldsymbol{\kappa}}]}.
\label{propagator1}
\end{equation}
The boundary points are indicated by $\overline{\mathbf{r}}_T=\overline{\mathbf{r}}(\beta)$ and
$\overline{\mathbf{r}}_0=\overline{\mathbf{r}}(0)$. To emphasize that the expression for the propagator is still a functional of the memory kernels and source functions, this dependence on $x,y$, and $\overline{\boldsymbol{\kappa}}$ is indicated in the square brackets.
The calculation of the propagator for $N=1$ has been performed in \cite{Adamowski}, and we largely base our derivation for the many-particle case in the rest of this section on the methods presented in \cite{Adamowski} and \cite{Brosens1997a}.

For a quadratic action functional given by expression (\ref{action1}), the path integral can be expanded around the classical paths that minimize the action functional to write
\begin{equation}
K_N[x, y, \overline{ \boldsymbol{\kappa}} ] \left( \overline{\mathbf{r}}_T  , \beta |  \overline{\mathbf{r}}_0, 0 \right) = K_N[x, y, \mathbf{0} ] \left( 0 , \beta |  0, 0 \right) e^{ -S_{\textrm{cl}} \left[ x,y, \overline{\boldsymbol{\kappa}} \right] \left( \overline{\mathbf{r}}_T, \overline{\mathbf{r}}_0\right)}.
\label{propagator2}
\end{equation}
Here, $S_{\textrm{cl}} \left[ x,y, \overline{\boldsymbol{\kappa}} \right] \left( \overline{\mathbf{r}}_T, \overline{\mathbf{r}}_0\right)$ is the action functional (\ref{action1}) evaluated along the classical paths that are found as solutions to the following set of integro-differential equations:
\begin{align}
&\ddot{\mathbf{R}}(\tau) - \int_{0}^{\beta} y(\tau-\sigma) \mathbf{R}(\sigma) d\sigma     + \mathbf{K}(\tau) = 0, \label{CMequation} \\
&\ddot{\mathbf{r}}_i(\tau) - \int_{0}^{\beta} x(\tau - \sigma) \mathbf{r}_i(\sigma) d\sigma   - \int_{0}^{\beta} \left[y(\tau-\sigma)- x(\tau-\sigma) \right] \mathbf{R}(\sigma) d\sigma + \boldsymbol{\kappa}_i(\tau) = 0. \label{rddot}
\end{align}
The center of mass coordinate $\mathbf{R}=\frac{1}{N} \sum_{i=1}^{N} \mathbf{r}_i$ decouples together with the center of mass source term $\mathbf{K}=\frac{1}{N} \sum_{i=1}^{N} \boldsymbol{\kappa}_i $ yielding an equation that has already been solved in \citep{Adamowski}.
Having obtained a solution to Eq.~(\ref{CMequation}), the last two terms in Eq.~(\ref{rddot}) can be seen as an effective source term, which allows us to solve Eq.~(\ref{rddot}) using the same approach. 
Substitution of the solutions into the action functional yields $S_{\textrm{cl}}(\overline{\mathbf{r}}_T, \overline{\mathbf{r}}_0)$, which can then be used to derive the fluctuation factor $K_N[x, y, \mathbf{0} ] \left( 0 , \beta |  0, 0 \right)$ in the same way as in \citep{Adamowski}.
This lengthy calculation can be somewhat shortened by writing the paths in terms of fluctuations around the center of mass, for which the derivation is presented in Appendix (\ref{Appendix A}).

As shown in Appendix (\ref{Appendix A}), the many-body propagator (\ref{propagator1}) factorizes in terms of single-particle propagators just as in the case of a harmonically coupled system \cite{Brosens1997a}:
\begin{equation}
K_N[x, y, \overline{\boldsymbol{\kappa}} ] \left( \overline{\mathbf{r}}_T  , \beta |  \overline{\mathbf{r}}_0, 0 \right) = \frac{ K[y, \sqrt{N} \mathbf{K} ] (\sqrt{N} \mathbf{R}_T , \beta | \sqrt{N} \mathbf{R}_0, 0 )}{K[x , \sqrt{N} \mathbf{K} ] (\sqrt{N} \mathbf{R}_T , \beta | \sqrt{N} \mathbf{R}_0, 0 )} \prod_{j=1}^{N} K[x , \boldsymbol{\kappa}_j] (\mathbf{r}_{j,T}, \beta | \mathbf{r}_{j,0}, 0 ).
\label{propagatorfactorization1}
\end{equation}
The propagators on the right-hand side of Eq.~(\ref{propagatorfactorization1}) are the single particle propagators for which the action functional (\ref{action1}) depends on a single memory kernel, making the notation of Eq.~(\ref{propagator1}) somewhat redundant. Hence, let us separately define the single particle propagator as a functional of only the memory kernel $x(\tau-\sigma)$:
\begin{equation}
K[x , \boldsymbol{\kappa}] (\mathbf{r}_T, \beta | \mathbf{r}_0, 0 ) = \int_{\mathbf{r}_0,0}^{\mathbf{r}_T,\beta} \mathcal{D} \mathbf{r} e^{-S^{(1)}[\mathbf{r},x,\boldsymbol{\kappa}]},
\end{equation}
where
\begin{equation}
S^{(1)}[\mathbf{r},x,\boldsymbol{\kappa}]=  \int_{0}^{ \beta} \frac{m \dot{\mathbf{r}}^2}{2} d\tau +  \frac{m}{2}  \int_0^{ \beta} d\tau \int_0^{\beta} d\sigma x(\tau - \sigma)  \mathbf{r}(\tau) \cdot \mathbf{r}(\sigma) -m  \int_0^{\beta} d \tau \mathbf{r} (\tau) \cdot \boldsymbol{\kappa}(\tau).
\end{equation}
In what follows, we will decompose the memory kernels and the source terms in their Fourier components $x_n$, $y_n$ and $\boldsymbol{\kappa}_n$, respectively, using the convention $f(\tau)= \sum_{n=-\infty}^{\infty} f_n e^{i \nu_n \tau}$, with $\nu_n = 2 \pi n / \beta$ being the bosonic Matsubara frequencies.
Following the method of \citep{Adamowski} and assuming the same stability conditions, we derive the following expression for the single-particle propagator with memory:
\begin{align}
K[x , \boldsymbol{\kappa}] (\mathbf{r}_T, \beta | \mathbf{r}_0, 0 ) &=   \left( \frac{m}{2 \pi \beta} \right)^{d/2} \left(  \frac{4}{\beta^3 x_0 \Delta_{x} } \right)^{d/2} \frac{1}{ {\displaystyle \prod_{k=1}  \left(  1 + \frac{\beta x_k}{\nu_k^2}\right)^{d} } }  \nonumber \\
& \times \exp \left[ -\frac{m}{2 \beta} A_x (\mathbf{r}_T - \mathbf{r}_0)^2 - \frac{m}{2\beta} \frac{1}{\Delta_{x}} (\mathbf{r}_T + \mathbf{r}_0)^2 
\nonumber \right. \\
&+  \frac{2m}{\beta} \frac{1}{\Delta_{x}}    \sum_{n } \frac{\boldsymbol{\kappa}_n  }{\nu_n^2 + \beta x_n} \cdot   (\mathbf{r}_T + \mathbf{r}_0) \nonumber \\
&- \frac{2 m }{\beta} \left( \frac{\beta}{2}    \sum_{n \neq 0}  \frac{ i \nu_n}{   \nu_n^2 + \beta x_n }  \boldsymbol{\kappa}_n \right)  \cdot  (\mathbf{r}_T - \mathbf{r}_0)  \nonumber  \\
&\left. -\frac{2 m }{\beta} \frac{1}{\Delta_{x}} \left( \sum_n \frac{\boldsymbol{\kappa}_{n}}{\nu_n^2 + \beta x_n}  \right)^2 + \frac{2m}{\beta} \left( \frac{\beta^2}{4}   \sum_n \frac{\boldsymbol{\kappa}_n \cdot \boldsymbol{\kappa}_{-n}}{\nu_n^2 + \beta x_n} \right) \right].
\label{SingleParticlePropagator}
\end{align}
In Eq.~(\ref{SingleParticlePropagator}) we have chosen a slightly different notation from that of \cite{Adamowski} to define the following dimensionless functionals of the memory kernel $x$:
\begin{align}
&A_x= \sum_{n=-\infty}^{\infty} \frac{\beta x_n}{\nu_n^2 + \beta x_n}, \label{Axn} \\
&\Delta_x=  \frac{4}{\beta^2} \sum_{n= -\infty}^{\infty} \frac{1}{\nu_n^2 + \beta x_n}. \label{Dxn}
\end{align}
In what follows we will generally assume $A_x>0$ and $\Delta_x>0$ to restrict ourselves to propagators (\ref{SingleParticlePropagator}) that are convergent for any combination of the boundary points.
Note that due to the previous assumption of $x_0 \neq 0$ and $y_0 \neq 0$ the functionals are well-defined when written in this form. 
Nevertheless, taking the limit $x_0,y_0 \rightarrow 0$ in the propagators still yields the appropriate expression, and this distinction will only become of importance in the partition function further on. 
 
  \section{Partition function for identical particles with memory} \label{Section 3}
  
The path-integral approach is naturally extended to the treatment of identical particles by taking all possible permutations of the end-points into account \cite{feynman1998statistical}. 
In this way, the canonical partition function for bosons is written as
\begin{equation}
\mathcal{Z}(N)=  \frac{1}{N!}  \sum_{P}  
\int d\overline{\mathbf{r}}
\int_{\overline{\mathbf{r}},0}^{P[\overline{\mathbf{r}}], \beta } \mathcal{D} 
\overline{\mathbf{r}}' ~ e^{-S^{(N)}[\overline{\mathbf{r}}', x , y , \overline{\boldsymbol{\kappa}} ]}.
\label{partitionfunction1}
\end{equation}
The path integral counts all possible paths from an ordered set of initial points $\overline{\mathbf{r}} = \lbrace \mathbf{r}_1 , \mathbf{r}_2 , ... \mathbf{r}_N \rbrace$ to a final set of points $P[\overline{\mathbf{r}}]=\lbrace P\mathbf{r}_1 , P\mathbf{r}_2 , ... P \mathbf{r}_N \rbrace$ where the coordinates are reordered by a permutation $P$ on a set of $N$, using the commonly used notation $P\mathbf{r}_1= \mathbf{r}_{P(1)}$. 
All possible values of the set $\overline{\mathbf{r}}$ are then integrated out, and the sum over all possible permutations $P$ is finally taken. 
The treatment can be straightforwardly extended to fermions by adding a factor $(-1)^P$ that provides a minus sign to all odd permutations.

The propagator (\ref{propagatorfactorization1}) exhibits the same factorization pattern as a harmonically coupled system of oscillators, and hence initially the approach of \citep{Brosens1997a} can be followed. 
The integration over all possible boundary points $\overline{\mathbf{r}}$ can be extended to include the center of mass variable through the introduction of a delta function,
\begin{equation}
   \int d\overline{\mathbf{r}} \rightarrow \int d\mathbf{R} \int 
   d\overline{\mathbf{r}} ~ \delta \left( \mathbf{R} - \frac{1}{N} \sum_{i} \mathbf{r}_i \right),
\end{equation} 
which is then written in its Fourier representation \cite{Brosens1997a}. 
This allows to separate the contribution of the center of mass propagators in Eq.~(\ref{propagatorfactorization1}) as follows:

\begin{equation}
\mathcal{Z}(N) = \frac{1}{(2 \pi)^3} \int d\mathbf{k} ~ \mathcal{Z}_R (N,\mathbf{k})  \mathcal{Z}_r(N,\mathbf{k}) , \label{kintegral}
\end{equation}
where

\begin{equation}
\mathcal{Z}_R (N,\mathbf{k})=  \int d\mathbf{R} ~ e^{i \mathbf{k} \cdot \mathbf{R}} \frac{ K[y, \boldsymbol{0} ] (\sqrt{N} \mathbf{R} , \beta | \sqrt{N} \mathbf{R}, 0 )}{K[x, \boldsymbol{0} ] (\sqrt{N} \mathbf{R} , \beta | \sqrt{N} \mathbf{R}, 0 )} \label{CMpartitionfunction}
\end{equation}
and
 \begin{equation}
 \mathcal{Z}_r(N,\mathbf{k})= \frac{1}{N!} \sum_P  \int d\overline{\mathbf{r}}
  \prod_{j=1}^{N} K[x , \boldsymbol{0}] ( P\mathbf{r}_j, \beta | \mathbf{r}_j, 0 ) e^{ - i \mathbf{k} \cdot  \mathbf{r}_j /N}. \label{rpartitionfunction}
\end{equation}

Note that we set the source functions $\boldsymbol{\kappa}_i=0$, as their main purpose was in deriving the fluctuation factor, and from now on we consider the action functional (\ref{action1}) without source terms. 
The integral in expression (\ref{CMpartitionfunction}) converges under the restriction $\Delta_x > \Delta_y$ and can be readily computed as a Gaussian integral, and calculating expression (\ref{rpartitionfunction}) will prove to be the main challenge. Following the standard approaches \cite{feynman1998statistical, Brosens1997a}, any permutation $P$ can be partitioned into $M_{\ell}$ disjoint permutation cycles of length $\ell$, which allows us to write:
 \begin{equation}
 \mathcal{Z}_r(N,\mathbf{k})=  \sum_{M_1,M_2,...,M_N}^{*}  \prod_{\ell =1}^{N} \frac{1}{ \ell^{M_\ell} (M_\ell)!} h_\ell (\mathbf{k})^{M_\ell},
 \label{rpartitionfunction3}
\end{equation}
where the $*$ symbol above the summation symbol indicates a constrained summation that has to obey $\sum_{\ell=1}^{N} \ell M_\ell=N$. In this representation the nested $N$-dimensional integral in expression (\ref{rpartitionfunction}) factorizes as a product of $\ell$-fold integrals that correspond to each permutation cycle:
\begin{equation}
h_\ell (\mathbf{k}) = \int d\mathbf{r}_1  ... \int d\mathbf{r}_\ell ~ K[x, \boldsymbol{0} ] ( \mathbf{r}_1, \beta | \mathbf{r}_\ell, 0 ) ... K[x, \boldsymbol{0} ] ( \mathbf{r}_3, \beta | \mathbf{r}_2, 0 )  K[x, \boldsymbol{0} ] ( \mathbf{r}_2, \beta | \mathbf{r}_1 0 ) e^{ - i\frac{1}{N}   \mathbf{k} \cdot  \sum_{j=1}^{\ell} \mathbf{r}_j}. \label{hq1}
\end{equation}
%
%
%

The next step is to obtain an expression for $h_\ell(\mathbf{k})$, which requires the computation of an $\ell$-dimensional integral in expression (\ref{hq1}). 
While high dimensional Gaussian integrals can always in principle be calculated by converting them into a linear algebra problem of finding a determinant of an $\ell$-dimensional matrix, finding an explicit expression for the latter is not always equally straightforward. 
In the approach of \citep{Brosens1997a}, which we have thus far followed very closely, the integral (\ref{hq1}) is calculated by relying on the composition property of the propagators. 
If the composition property holds, then $h_\ell(\mathbf{k})$ becomes the single-particle partition function of exactly the same system as described by the single-particle propagator, but at an inverse temperature $\ell \beta$ and with additional delta-kicks to account for the $\mathbf{k}$-exponent. This partition function can then be readily computed with standard path integration methods. 
This trick is not applicable here, as the propagator with memory (\ref{SingleParticlePropagator}), does not obey the composition property. 
This can be easily seen by noting that the action functional (\ref{action1}) can not just be split into a sum of two parts on respective time intervals. 
In Appendix (\ref{Appendix B}) we show how integral (\ref{hq1}) can be directly computed and obtain the following result in $d$ dimensions:
\begin{equation}
h_\ell (\mathbf{k})= Q_x^{\ell d} \frac{1}{ \left| 2 \sinh( \frac{ \ell }{2}  \textrm{arccosh}\left[ \frac{A_x \Delta_{x} + 1}{A_x \Delta_{x} - 1} \right] ) \right|^d}  \exp \left( -\frac{ \ell k^2 \beta }{ 8 N^2 m} \Delta_{x}   \right) \label{hqmain}, 
\end{equation}
where
\begin{equation}
Q_x=\frac{1}{ \prod_{k=1}  \left(  1 + \frac{\beta x_k}{\nu_k^2}\right) }   \left( \frac{1}{\beta^3 x_0}  \frac{ 4}{\left| A_x \Delta_{x} - 1 \right|} \right)^{1/2}.
\label{Qxn}
\end{equation}
The functional form of $h_\ell(\mathbf{k})$ is very similar to that found in \citep{Brosens1997a}. 
The main differences are that the oscillator-frequency dependent parts are now replaced by expressions containing $\Delta_{x}$ and $A_x$, functionals of the memory kernel, appearing in the argument of the hyperbolic sine and the exponential. 
An additional factor $Q_x$ appears, which equals 1 when the memory kernel $x$ corresponds to a harmonic oscillator without memory. 

The choice of writing expression (\ref{hqmain}) in terms of the hyperbolic sine has the advantage of being maximally illustrative in regard to how changes due to memory arise on top of previously known expressions in \cite{Brosens1997a}. However, due to this choice some particular care should be taken when $\Delta_x A_x <1$. In this case each of the two factors in the determinant (\ref{determinant2}) in Appendix \ref{Appendix B} can become negative, and the complex modulus should be added after taking the square root if the factors are to be separated as in (\ref{hqmain}) and (\ref{Qxn}). For simple harmonic oscillator systems, and the model system considered in Section \ref{Section 4}, $\Delta_x A_x>1$ and this subtlety can be safely ignored.

The expression for the partition function (\ref{kintegral}) can now be computed. 
The center of mass $\mathcal{Z}_R(N,\mathbf{k})$ can be calculated from the propagators, and now that the $\mathbf{k}$-dependence of $h_\ell(\mathbf{k})$ is known, the $\mathbf{k}$-integral in (\ref{kintegral})  can be performed. After some algebraic work, one obtains:
\begin{align}
\mathcal{Z}(N) = \mathbb{Z}(N) Q_x^{Nd} \left( \frac{\beta x_0}{\beta y_0} \right)^{d/2} \prod_{k=1}^{\infty} \left( \frac{ 1 + \frac{\beta x_k}{\nu_k^2}}{1+ \frac{\beta y_k}{\nu_k^2}} \right)^{d} \hspace{3pt}  .
\label{partitionfunction4}
\end{align}
 with $\mathbb{Z}$ given by:
\begin{equation}
\mathbb{Z}(N) =  \sum_{M_1,M_2,...,M_N}^{*}  \prod_{\ell=1}^{N} \frac{1}{ \ell^{M_\ell} (M_\ell)!} \frac{1}{ \left|  2 \sinh( \frac{ \ell }{2}  \textrm{arccosh}\left[ \frac{A_x \Delta_{x} + 1}{A_x \Delta_{x} - 1} \right] ) \right|^{M_\ell d}}.
 \label{Ztilde}
\end{equation}
Note that due to the presence of the additional factor in expression (\ref{partitionfunction4}) it is now the product $Q_x^{Nd} \mathbb{Z}(N)$ that represents the partition function in the absence of two-body interactions, extending the result of \cite{Brosens1997a}. Following the approach in \cite{Brosens1997a}, the constrained summation (\ref{Ztilde}) can be transformed into a recurrence relation:
\begin{equation}
\mathbb{Z}(N) =  \frac{1}{N} \sum_{k=0}^{N-1} \mathbb{Z}(k) \left| 2 \sinh\left[ \frac{(N-k)}{2} \text{arccosh}\left( \frac{A_x \Delta_{x} + 1}{A_x \Delta_{x} - 1} \right) \right] \right| ^{-d}.
\label{recurrence1}
\end{equation}
The recurrence relation requires an initial value, and it can be seen that $\mathbb{Z}(0)=1$ yields the correct $\mathcal{Z}(1)$ result according to expression (\ref{partitionfunction4}). 
Alternatively, the factor $Q_x$ could be absorbed in the definition of $\mathbb{Z}(N)$, but then the recurrence would have to start from $\mathbb{Z}(0)=Q_x^{Nd}$.

As a consistency check, consider the specific choice $x(\tau-\sigma)= w^2 \delta(\tau-\sigma)$ and $y(\tau-\sigma)=\Omega^2 \delta(\tau-\sigma)$ for which the action functional (\ref{action1}) exactly corresponds with the system of coupled oscillators in \cite{Brosens1997a}. 
The different Matsubara sums and products in Eq.~(\ref{partitionfunction4}) can now be readily computed to find:
\begin{align}
&\left( \frac{\beta x_0}{\beta y_0} \right)^{d/2} \prod_{k=1}^{\infty} \left( \frac{ 1 + \frac{\beta x_k}{\nu_k^2}}{1+ \frac{\beta y_k}{\nu_k^2}} \right)^{d} = \frac{\sinh ( \frac{\beta w}{2} )^d}{\sinh( \frac{\beta \Omega}{2} )^d},
\end{align}
and $Q_x=1$. In particular, the resulting hyperbolic cosine from
\begin{align}
  \frac{A_x \Delta_{x} + 1}{A_x \Delta_{x} - 1} &= \cosh( \beta w ) 
\end{align}
allows to cancel the inverse hyperbolic cosine in the weight factor of the recurrence relation (\ref{recurrence1}). 
Substituting these results, the expression for the partition function in \cite{Brosens1997a} is exactly retrieved in this limit. 

\section{Example application: open quantum system of identical oscillators} \label{Section 4}

\begin{figure}
\includegraphics[width=0.7\columnwidth]{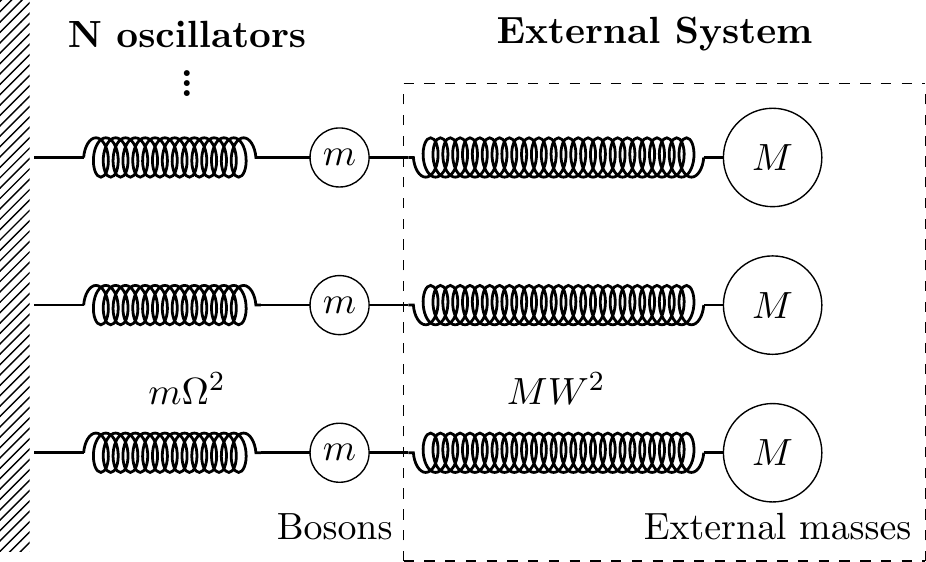}
\caption{A one dimensional depiction of the system described by (\ref{Ltot}).} 
\label{Figure0}
\end{figure}

In this section we present a brief example application of the derived expressions to a stylized model of an open quantum system of identical particles.
We consider the system depicted in Fig.~\ref{Figure0} of non-interacting bosons in a harmonic trap with frequency $\Omega$, coupled to an environment. 
The effect of the environment is modeled as a harmonic coupling with frequency $W$ of each boson to a fictitious particle with mass $M$. This model corresponds to a particular equal particle case of the more general models studied in \cite{Hasegawa1,Hasegawa2} for distinguishable particles of the system, which we will here consider for bosons. Note that because of the Bose statistics that have to be imposed, this model is more than simply $N$ unrelated copies of a two-particle system.

The fictitious particles are taken to be uncoupled and distinguishable, which could represent an environment with a far slower relaxation than the bosonic system. The (Euclidean) Lagrangian of the full system corresponding to the partition function $\mathcal{Z}_{\textrm{tot}}$ is therefore  given by

\begin{equation}
    L_{\textrm{tot}} = \sum_{i=1}^{N} \left(  \frac{m}{2} \dot{\mathbf{r}}_i^2 +  \frac{m \Omega^2}{2} \mathbf{r}_i^2+\frac{M}{2} \dot{\mathbf{Q}}_i^2 + \frac{MW^2}{2} \left( \mathbf{r}_i - \mathbf{Q}_i \right)^2 \right), 
    \label{Ltot}
\end{equation}
and the Lagrangian of the external system is defined as
\begin{equation}
L_{\textrm{f}}= \sum_{i=1}^{N} \left( \frac{M}{2} \dot{\mathbf{Q}}_i^2 + \frac{MW^2}{2}  \mathbf{Q}_i ^2 \right).
\label{Lf}
\end{equation}
It is important to note that through (\ref{Lf}) we adopt the view of the external system as defined in \cite{Ingold2009}. The external system is considered to be the whole of the fictitious particles and springs with constant $MW^2$ that are attached to the degrees of freedom of interest.
The fictitious particles can be integrated out as in \cite{Tempere2009} to obtain the reduced partition function:

\begin{equation}
    \mathcal{Z}(N) = \frac{\mathcal{Z}_{\textrm{tot}}(N)}{\mathcal{Z}_{\textrm{f}}(N)},
    \label{dividedZf}
\end{equation}
where $\mathcal{Z}_{\textrm{f}}$ is the partition function of the external system corresponding to (\ref{Lf}).
Expression (\ref{dividedZf}) is exactly the identical particle extension of one of the stylized models of an open quantum system considered in \cite{Ingold2009}.
The resulting $\mathcal{Z}(N)$ can now be cast in the form of (\ref{partitionfunction1}), where the memory kernels in the action functional (\ref{action1}) are given by

\begin{equation}
x(\tau-\sigma)=y(\tau-\sigma) = \frac{MW^2}{m} \left[  \frac{W^2+ \frac{m}{M} \Omega^2}{W^2} \delta(\tau-\sigma) - \frac{W \cosh( W \left[ |\tau-\sigma|-\beta/2 \right])}{2 \sinh(W \beta /2)}  \right] . \label{memorykernel1}
\end{equation}
This is the simplest translationally non-invariant model that provides a memory kernel $x(\tau-\sigma)$ with non-trivial memory effects for the recurrence relation (\ref{recurrence1}). 
The functionals $A_x$ and $\Delta_x$ are obtained after computing the Matsubara summations in expressions (\ref{Axn}) and (\ref{Dxn}):

\begin{align}
&A_x   =\frac{\beta \omega_+}{2} \coth( \frac{\beta \omega_+}{2} ) \gamma_+ + \frac{\beta \omega_-}{2} \coth( \frac{\beta \omega_-}{2} ) \gamma_- , \label{Axn1} \\
&\Delta_x =   \frac{2}{\beta \omega_+} \coth( \frac{\beta \omega_+}{2} )  \gamma_+  + \frac{2}{\beta \omega_-} \coth( \frac{\beta \omega_-}{2} )    \gamma_- , \label{Dxn1}
 \end{align}
where

\begin{align}
  &\omega_{\pm}^2 =  \frac{\frac{m+M}{m}  W^2 +    \Omega^2 \pm   \sqrt{ \left( \frac{m+M}{m} W^2 + \Omega^2 \right)^2 - 4 W^2 \Omega^2} }{2}  , \\
  &\gamma_{\pm} = \frac{1}{2} \left[ 1 \pm \frac{\Omega^2 + \left( \frac{M}{m}-1 \right)W^2}{\omega_+^2-\omega_-^2} \right].
\end{align}
The frequencies $\omega_{\pm}$ that diagonalize the full system \cite{Ingold2009} therefore naturally appear in the calculation. 

Since for this system $y_n=x_n$, the interaction factor in front of (\ref{partitionfunction4}) cancels out and the partition function $\mathcal{Z}(N)$ is written as a product of only two factors, $Q_x^{Nd}$ and the recurrence part $\mathbb{Z}(N)$. The Matsubara product in (\ref{Qxn}) can be computed for the specific memory kernel (\ref{memorykernel1}), which allows us to write
\begin{equation}
Q_x =\frac{2 \sinh \left( \frac{\beta W}{2} \right)}{\beta W}   \frac{\beta \omega_+}{2 \sinh \left( \frac{\beta \omega_+}{2} \right)} \frac{\beta \omega_-}{2 \sinh \left( \frac{\beta \omega_-}{2} \right)}  \left( \frac{1}{\beta^3 x_0}  \frac{ 4}{ \Delta_x  A_x - 1} \right)^{1/2},
\label{Qxn2}
\end{equation}
with $\Delta_x$ and $A_x$ known from (\ref{Dxn1}) and (\ref{Axn1}). In three dimensions the recurrence relation (\ref{recurrence1}) for $\mathbb{Z}(N)$ has no known solution, and has to be computed numerically. As shown in the approach of \cite{Brosens1997a} a numerically stable implementation is obtained by defining
\begin{equation}
    b= e^{-q} , \hspace{30pt}
q=\text{arccosh}\left[ \frac{\Delta_x A_x + 1}{\Delta_x A_x - 1} \right], \label{q}
\end{equation}
and without loss of generality proposing the following way of writing the recurrence factor:
\begin{equation}
\mathbb{Z}(N) =  \prod_{j=1}^{N} \rho_j \frac{b^{ \frac{3}{2}}}{ \left( 1 - b^{ j} \right)^{3}}.
\label{recurrence2}
\end{equation} 
This fixes the first coefficient $\rho_1=1$, and after substitution of (\ref{recurrence2}) into (\ref{recurrence1}) a recurrence relation for $\rho_N$ is found:
\begin{equation}
\rho_N = \frac{1}{N} \frac{ \left(1- b^N \right)^3}{\left(1 - b \right)^3} \left[ 1  + \sum_{k=0}^{N-2} \frac{(1-b)^3}{\left( 1 - b^{ (N - k)} \right)^{3}}  \prod_{j=k+1}^{N-1} \frac{(1-b^j)^3}{\rho_j} \right].
\label{rhoN}
\end{equation}
Due to the additional factor in the expression for the partition function $\mathcal{Z}(N)=Q_x^{Nd} \mathbb{Z}(N)$, the internal energy and specific heat of the system are written as a sum of two terms:
\begin{align}
    &U(N)= U_Q(N) + \mathbb{U}(N) = - 3N \partial_\beta \log( Q_x) - \partial_\beta  \log( \mathbb{Z} ) , \label{UN} \\
    &C(N)=C_Q(N) + \mathbb{C}(N)= 3N k_B \beta^2 \partial^2_\beta \log( Q_x) + k_B \beta^2 \partial^2_\beta  \log( \mathbb{Z} ) \label{CN}.
\end{align}
Analytical expressions for $U_Q$ and $C_Q$ can straightforwardly be calculated from the factor $Q_x$ in (\ref{Qxn2}). 
The recurrence relations for $\mathbb{U}(N)$ and $\mathbb{C}(N)$ are obtained after computing the partial derivatives of $\log(\mathbb{Z})$ by combining (\ref{recurrence2}) with (\ref{rhoN}):

\begin{align}
\frac{\mathbb{U}(N)}{ \partial_\beta q}  =& \frac{1}{N} \frac{1}{\rho_N} \frac{(1-b^N)^3}{(1-b)^3}  \left(   \frac{\mathbb{U}(N-1) }{ \partial_\beta q}  + \frac{3}{2} \frac{1+b}{1-b} \right. \nonumber \\
 + & \left. \sum_{k=0}^{N-2} \frac{(1-b)^3}{\left( 1 - b^{ (N - k)} \right)^{3}}   \left[ \frac{\mathbb{U}(k) }{ \partial_\beta q}   +   \frac{3(N-k)}{2} \frac{1+b^{(N-k)}}{1-b^{(N-k)}}  \right]\prod_{j=k+1}^{N-1} \frac{ (1-b^j)^3  }{ \rho_j} \right),
\end{align}
and
\begin{align}
&\mathbb{C}(N) k_B^{-1} = \frac{1}{N} \sum_{k=0}^{N-1} \frac{1}{(1-b^{(N-k)})^3}\prod_{j=k+1}^{N} \frac{ (1-b^j)^3  }{  \rho_j}\left( k_B^{-1} \mathbb{C}(k)  \vphantom{ \frac{1}{1} }\right.  \nonumber \\
&+\beta^2\left[   \frac{3(N-k)}{2} \frac{1+b^{(N-k)}}{1-b^{(N-k)}}  \partial_\beta q  +  \mathbb{U}(k)- \mathbb{U}(N)   \right] \left[   \frac{3(N-k)}{2} \frac{1+b^{(N-k)}}{1-b^{(N-k)}}  \partial_\beta q  +  \mathbb{U}(k)   \right]\nonumber \\ 
 & \left.        +\beta^2    3(N-k)^2   \frac{ b^{(N-k)} }{(1-b^{(N-k)})^2} (\partial_\beta q)^2 - \beta^2    \frac{3(N-k)}{2}  \frac{1+b^{(N-k)}}{1-b^{(N-k)}} \partial^2_\beta q  \right).
\end{align}
Here, the recurrence formulas are initiated from $\mathbb{U}(0)=0$ and $\mathbb{C}(0)=0$, and the partial derivatives $\partial_\beta q$ and $\partial^2_\beta q$ can be analytically computed from (\ref{q}) since $\Delta$ and $A$ are known. 
\begin{figure}
\includegraphics[width=0.6\columnwidth]{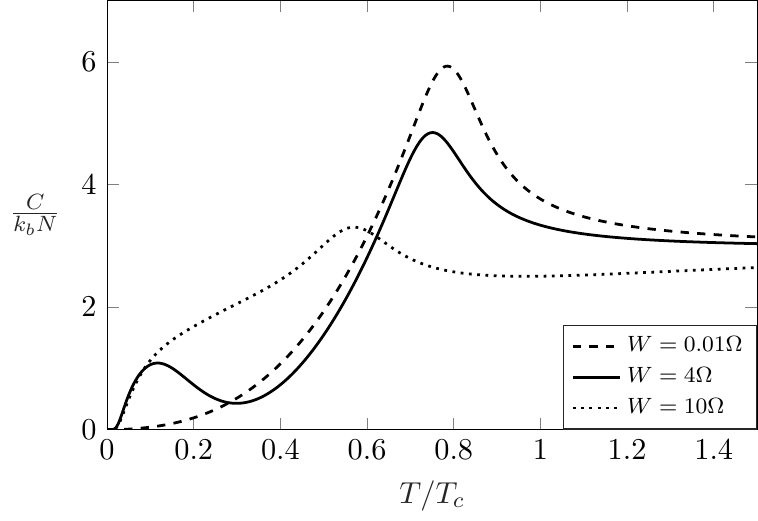}
\caption{ Specific heat per particle as a function of the temperature of $N=100$ non-interacting bosons in a harmonic potential, harmonically coupled to external masses for $M=m$. The results are shown for three coupling strengths $W=[0.01\Omega, 4\Omega, 10\Omega]$ plotted by respectively the dashed, solid and dotted lines.} 
\label{Figure1}
\end{figure}

The specific heat (\ref{CN}) is shown in Fig.~\ref{Figure1} as a function of the temperature, measured with respect to the critical temperature in the absence of the external system 
$k_B T_c= \hbar \Omega \left( N / \zeta(3)\right)^{1/3} $, 
with $\zeta(x)$ being the Riemann zeta function. We can clearly observe the main bosonic condensation peak slightly below the critical temperature, which at weak coupling corresponds exactly to the result in \citep{Brosens1997a}. The sharpness of the peak fades towards stronger coupling with the external system but nevertheless remains visibly present.
In addition to the main condensation peak, at an intermediate coupling strength an anomalous dip and peak are observed at low temperatures. 
These anomalous features in the specific heat of open quantum systems have been studied for distinguishable particles in \cite{Ingold2009,Hasegawa2,Ingold2012}, where it is shown that the specific heat can even become negative for certain systems. 
This is explained in \cite{Ingold2009,Hasegawa2} by the fact that the specific heat (\ref{CN}) is the difference of the specific heats of the system and the trapped fictitious particles as defined in the partition function (\ref{dividedZf}), and a more extensive interpretation can be found in \cite{Ingold2012}. 

We can also note that the high and low temperature limits of the specific heat are in agreement with \cite{Hasegawa2}. From expression (\ref{Qxn2}) we can see that at high temperatures for $\beta \rightarrow 0$, $Q_x$ approaches a finite value and hence the first part of the specific heat $C_Q(N)$ in (\ref{CN}) goes to zero. In the same limit the recurrence part of the partition function can be shown to diverge as $\mathbb{Z} \sim \beta^{-Nd}$, from which follows $C(N)=3Nk_B$. In the low-temperature limit $\beta \rightarrow \infty$ one can show that in the presence of the environment $\mathbb{Z}$ remains finite, and $Q_x$ becomes an exponential function of $\beta$, from which follows $C(N)=0$.

\begin{figure}[!htbp]
\includegraphics[width=0.8\columnwidth]{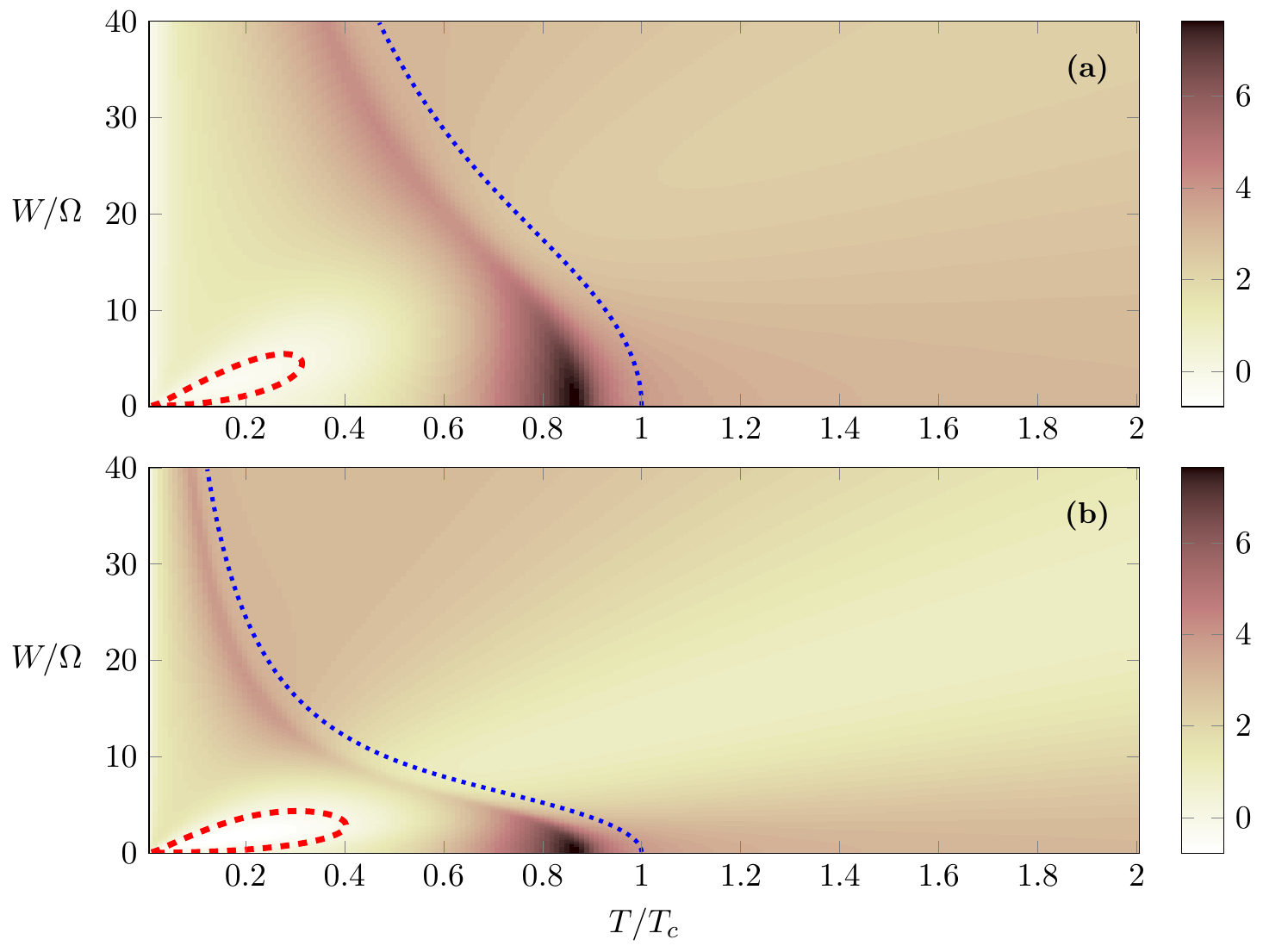}
\caption{ Color map of the specific heat per particle $C/(k_B N)$ for $N=500$ bosons, for (a) $M=m$ and (b) $M=10m$. The dashed loop in the bottom left corner of each color map indicates the region where the specific heat becomes negative. The dotted line indicates the effective temperature obtained from (\ref{effectivetemp}).} 
\label{Figure2}
\end{figure}

An overview of the structure of the main condensation peak and the anomalous dip is presented in Fig.~\ref{Figure2}. For both a light and a heavy mass $M$ of the fictitious particles, remnants of the bosonic condensation peak remain visible up to strong coupling with the external system. At low temperatures and weaker coupling the anomalous dip can be seen as region of lighter shading. In contrast to the single-particle case for this system \cite{Ingold2009}, we find that the anomalous dip can drop below zero for bosons in Fig.~\ref{Figure2}, where the dashed loop indicates a region of negative specific heat. 

As can also be seen from Fig.~\ref{Figure2}, coupling with the external system significantly lowers the effective critical temperature of the bosons. This can be understood by noting that the generalized bosonic recurrence relation (\ref{recurrence1}) is nothing else than the recurrence relation for harmonically trapped bosons where the trap frequency is replaced by a temperature-dependent quantity:

\begin{equation}
\tilde{\Omega}(T) = \frac{1}{\beta}  \text{arccosh}\left[ \frac{\Delta_x A_x+ 1}{\Delta_x A_x - 1} \right],
\end{equation}
which allows to define an effective critical temperature as the solution of:
\begin{equation}
\frac{\tilde{T}_c}{T_c} = \frac{\tilde{\Omega}(\tilde{T}_c)}{\Omega} .
\label{effectivetemp}
\end{equation}
The results are plotted as the dotted lines in Fig.~\ref{Figure2} and agree well with the behavior of the condensation peak. 
It is important to note that only the recurrence part is correctly reproduced by substituting $\Omega \rightarrow \tilde{\Omega}(T)$ in the harmonic oscillator result. 
The factor $Q_x$ in front of the partition function (\ref{partitionfunction4}) is not retrieved this way because it is entirely absent in the harmonic case. As the latter however can be taken out of the recurrence relation it is no surprise that it should play no significant role in the inherently bosonic features of the system, and the behavior of the condensation peak is accurately reproduced by (\ref{effectivetemp}).

\section{Conclusion}

In this work we presented an approach that incorporates the effects of retarded interactions in the path integral formalism for identical particles.  
First, the many-body propagator for distinguishable particles was derived and shown to exhibit the same factorization pattern in terms of single-particle propagators as seen in harmonically coupled systems without retardation \citep{Brosens1997a}. 
However, the main difference is that the single-particle propagators no longer obey the composition property when the system has memory. 
This complicates the computation of a class of integrals appearing in the derivation of the partition function, for which we obtain explicit expressions by utilizing the properties of circulant matrices. 
The resulting expression for the partition function is a functional applicable to a general class of memory kernels, and is shown to reduce to the known result for harmonically coupled systems without memory in the appropriate limit.

The results were then applied to study the specific heat of non-interacting bosons in a harmonic trap coupled to an external system of fictitious masses. 
This provides the simplest model system that yields non-trivial memory effects in the condensation recurrence relation. 
We show that the presence of the environment shifts the bosonic condensation to lower temperatures and significantly smooths out the Bose condensation peak in the specific heat, which nevertheless remains visible even at strong coupling. To better understand these types of open systems, and in particular to calculate the density and the pair correlation function, expressions for the identical particle one-and two-point generating functionals are required. The results presented here pave the way to compute these quantities. These will in turn allow  one to study the autocorrelation functions, occupation numbers, and formulate the most general harmonic variational approach for identical particles.

\begin{acknowledgments}
We gratefully acknowledge fruitful discussions with F. Brosens, S.N. Klimin and M. Houtput. 
T.I. acknowledges the support of the Research Foundation-Flanders (FWO-Vlaanderen) 
through the PhD Fellowship Fundamental Research, Project No. 1135521N. 
We also acknowledge financial support from Research Foundation-Flanders (FWO-Vlaanderen) Grant No. G.0618.20.N, and  from the research council of the University of Antwerp.

\end{acknowledgments}

\appendix

\section{Derivation of the distinguishable particle propagator}\label{Appendix A}

For the single-particle limit of (\ref{action1}), the classical action is calculated in \citep{Adamowski}. 
For completeness and due to slightly different notations, we briefly summarize the calculation below. Consider the single-particle action functional:
\begin{equation}
S^{(1)}[\mathbf{r},x,\boldsymbol{\kappa}]=  \int_{0}^{ \beta} \frac{m \dot{\mathbf{r}}^2}{2} d\tau +  \frac{m}{2}  \int_0^{ \beta} d\tau \int_0^{\beta} d\sigma x(\tau - \sigma)  \mathbf{r}(\tau) \cdot \mathbf{r}(\sigma) -m  \int_0^{\beta} d \tau \mathbf{r} (\tau) \cdot \boldsymbol{\kappa}(\tau).
\label{appendixaction1}
\end{equation}
The classical path is found as the solution to the following integro-differential equation with boundary conditions $\mathbf{r}_T  = \mathbf{r}(\beta)$ and $\mathbf{r}_0 = \mathbf{r}(0)$:
\begin{equation}
\ddot{\mathbf{r}}(\tau) - \int_{0}^{\beta} x(\tau - \sigma) \mathbf{r}(\sigma) d\sigma   + \boldsymbol{\kappa}(\tau)= 0.
\label{appendixrddot}
\end{equation}
In \citep{Adamowski}, the following Fourier decomposition is proposed:
\begin{equation}
\mathbf{r}_{\textrm{cl}}(\tau)= \mathbf{r}_0 + (\mathbf{r}_T - \mathbf{r}_0) \frac{\tau}{\beta}- \frac{\mathbf{A}_0}{2}  \tau(\tau-\beta) + \sum_{n \neq 0 } \frac{\mathbf{A}_n}{\nu_n^2} \left( e^{i \nu_n \tau} -1 \right),
\label{ansatz1}
\end{equation} 
where after substitution into (\ref{appendixrddot}), the following solutions are found (assuming $x_0 \neq 0$, otherwise the appropriate limit should be taken):

\begin{align}
&\mathbf{A}_0   =  \frac{4}{\beta^2 \Delta_x} \left( \sum_{n } \frac{ \boldsymbol{\kappa}_n}{ \nu_n^2 + \beta x_n} - \frac{1}{2} (\mathbf{r}_T + \mathbf{r}_0) \right) ,   \label{A0} \\
 &\mathbf{A}_n =  \frac{\beta x_n}{\nu_n^2 + \beta x_n} \mathbf{A}_0 + \frac{1}{\left( 1 + \frac{\beta x_n}{\nu_n^2} \right) } \left( \boldsymbol{\kappa}_n + x_n \frac{ \mathbf{r}_T - \mathbf{r}_0}{i \nu_n }  \right) . \label{An}
\end{align}
The coefficients can be substituted into (\ref{ansatz1}) to obtain an explicit expression for the classical solution $\mathbf{r}_{\textrm{cl}}(\tau)$ and its Fourier components $\mathbf{r}_n$. 
After integrating the kinetic energy by parts, and writing the remaining source term integral in Fourier space, the classical action can be written as

\begin{equation}
S_{\textrm{cl}}^{(1)}[x,\boldsymbol{\kappa}](\mathbf{r}_T, \mathbf{r}_0)= \frac{m}{2} \left( \dot{\mathbf{r}}_{\textrm{cl}}(\beta) \cdot \mathbf{r}_T - \dot{\mathbf{r}}_{\textrm{cl}}(0) \cdot  \mathbf{r}_0 \right)  -\frac{m \beta}{2} \sum_n \mathbf{r}_n  \cdot \boldsymbol{\kappa}_{-n} .
\label{classaction1}
\end{equation}
By taking the derivative of (\ref{ansatz1}) and substituting its boundary points to find the first part, and performing the Fourier sum using $\mathbf{r}_n= \frac{\boldsymbol{\kappa}_n - \mathbf{A}_n}{\beta x_n}$ to find the second part, the single-particle classical action becomes
\begin{align}
S^{(1)}_{\textrm{cl}}[x,\boldsymbol{\kappa}](\mathbf{r}_T, \mathbf{r}_0) =& \frac{m}{2 \beta} A_x (\mathbf{r}_T - \mathbf{r}_0)^2 + \frac{m}{2\beta} \frac{1}{\Delta_x} (\mathbf{r}_T + \mathbf{r}_0)^2 \nonumber \\
 -&  \frac{2m}{\beta }  \frac{1}{\Delta_x} \left( \mathbf{r}_T + \mathbf{r}_0 \right)  \cdot \sum_n \frac{\boldsymbol{\kappa}_n }{\nu_n^2+ \beta x_n}  + \frac{2m}{\beta} \left( \mathbf{r}_T - \mathbf{r}_0 \right) \cdot \left( \frac{\beta}{2} \sum_n  \frac{ i \nu_n \boldsymbol{\kappa}_n}{\nu_n^2+ \beta x_n} \right)  \nonumber \\
+&   \frac{2 m }{\beta} \frac{1}{ \Delta_x} \left( \sum_n \frac{\boldsymbol{\kappa}_n}{\nu_n^2 + \beta x_n}\right)^2 - \frac{2m }{ \beta} \left( \frac{\beta^2}{4} \sum_n \frac{\boldsymbol{\kappa}_n \cdot \boldsymbol{\kappa}_{-n}}{\nu_n^2 + \beta x_n} \right).
\label{classaction2}
\end{align}

For the source terms, some care should be taken regarding pointwise convergence when performing calculations in Fourier space, as pointed out in \cite{Adamowski}. 
For example, when considering a source function $\boldsymbol{\kappa}(\tau)= f \delta(\tau-\sigma)$  for $\sigma = 0$ or $\sigma=\beta$, the correct result should be derived by considering $\sigma \in ]0,\beta[$ and respectively taking the limit of $\sigma \rightarrow 0^+$ or $\sigma \rightarrow \beta^-$ rather than direct substitution due to discontinuities at the edge. 
Taking care of the appropriate limits, the known results for e.g. the harmonic oscillator or the kicked particle are readily obtained from (\ref{classaction2}).

To obtain the many-particle extension of this result for the action functional (\ref{action1}), a similar but lengthier calculation was performed starting from equations (\ref{CMequation}) and (\ref{rddot}) by first finding $\mathbf{R}_{\textrm{cl}}(\tau)$ with the previous method and then using this result to solve the equation for $\mathbf{r}_{\textrm{cl}}^{(i)}(\tau)$.
However, in line with \cite{Brosens1997a}, a somewhat shorter argument yielding the same result can be formulated by switching to the variable $\mathbf{u}_{i}= \mathbf{r}_i - \mathbf{R}$ at the level of the classical equations:
\begin{align}
&\ddot{\mathbf{R}}(\tau) - \int_{0}^{\beta} y(\tau-\sigma) \mathbf{R}(\sigma) d\sigma     + \mathbf{K}(\tau) = 0, \label{CMequation11} \\
&\ddot{\mathbf{u}}_i(\tau) - \int_{0}^{\beta} x(\tau - \sigma) \mathbf{u}_i(\sigma) d\sigma   + \boldsymbol{\kappa}_i(\tau) - \mathbf{K}(\tau) = 0, \label{uddot}
\end{align} 
with boundary conditions $\mathbf{u}_{i,(T,0)}= \mathbf{r}_{i,(T,0)} - \mathbf{R}_{(T,0)}$. In addition, the solution is subject to the constraint $\sum_{i} \mathbf{u}_i (\tau)= \mathbf{0}$. 
The many-body classical action corresponding to (\ref{action1}), written in terms of the coordinates $\mathbf{u}_i$ and $\mathbf{R}$ yields:
\begin{equation}
S_{\textrm{cl}} \left[ x,y, \overline{\boldsymbol{\kappa}}\right] \left( \overline{\mathbf{r}}_T, \overline{\mathbf{r}}_0 \right) = \sum_{i=1}^{N} S^{(1)}_{\textrm{cl}} \left[ x,\boldsymbol{\kappa}_i-\mathbf{K} \right] \left( \mathbf{u}_{i,T}, \mathbf{u}_{i,0} \right) +  S^{(1)}_{\textrm{cl}}\left[ y, \sqrt{N} \mathbf{K} \right] \left( \sqrt{N} \mathbf{R}_T, \sqrt{N} \mathbf{R}_0 \right).
\label{classaction3}
\end{equation}
Here, we have used the property $\sum_{i} \mathbf{u}_i (\tau)= \mathbf{0}$ to drop a number of terms, and add an additional source term in $\mathbf{K}(\tau)$ to obtain the difference of source terms $\boldsymbol{\kappa}-\mathbf{K}$ in the first term of (\ref{classaction3}). 
Through direct substitution of the boundary conditions $\mathbf{u}_{i,(T,0)}= \mathbf{r}_{i,(T,0)} - \mathbf{R}_{(T,0)}$ and source term $\boldsymbol{\kappa}_i-\mathbf{K}$ into (\ref{classaction2}), one can easily confirm that
\begin{align}
S_{\textrm{cl}} \left[ x,y, \overline{\boldsymbol{\kappa}}\right] \left( \overline{\mathbf{r}}_T, \overline{\mathbf{r}}_0 \right) =  \sum_{i=1}^{N} S^{(1)}_{\textrm{cl}}\left[ x , \boldsymbol{\kappa}_i\right]  \left( \mathbf{r}_{i,T}, \mathbf{r}_{i,0} \right) &+  S^{(1)}_{\textrm{cl}} \left[ y, \sqrt{N} \mathbf{K} \right] \left( \sqrt{N} \mathbf{R}_T, \sqrt{N} \mathbf{R}_0 \right) \nonumber \\
&- S^{(1)}_{\textrm{cl}} \left[ x, \sqrt{N} \mathbf{K} \right] \left(\sqrt{N} \mathbf{R}_T, \sqrt{N} \mathbf{R}_0 \right) .
\label{classicalaction4}
\end{align}

Next, we have to find the fluctuation factor of the propagator $ K_N[ x,  y, \boldsymbol{0} ]  (0 , \beta | 0, 0 )$ as defined in section (\ref{Section 2}). 
While the decomposition of the classical action (\ref{classicalaction4}) strongly suggests a similar factorization for the fluctuation factor, let us present a complete overview of the calculation.  
Following the approach in \cite{Adamowski}, we consider the many-particle fluctuation factor $K_N[\lambda x, \lambda y,  \overline{\boldsymbol{0}}] \left( 0 , \beta | 0, 0 \right)$ where the memory kernels are scaled by a variable $\lambda$, and define
\begin{equation}
J(\lambda)= \log \left(  K_N[ \lambda x,  \lambda y, \boldsymbol{0} ]  (0 , \beta | 0, 0 ) \right).
\end{equation}
The logarithm of the fluctuation factor $J(1)$ can then be written as
\begin{equation}
J(1)=J(0) + \int_0^1 d\lambda \frac{\partial J(\lambda)}{\partial \lambda} = J(0) + \int_0^1 d\lambda \frac{\frac{\partial}{\partial \lambda}  K_N[  \lambda x,  \lambda y, \boldsymbol{0} ]  (0 , \beta | 0, 0 )}
{ K_N[  \lambda x,  \lambda y, \boldsymbol{0} ]  (0 , \beta | 0, 0 )} ,
\label{J1}
\end{equation}
where of course $J(0)= \frac{Nd}{2} \log( \frac{m}{2\pi \beta} )$ is the known free-particle result in $d$ dimensions. In path-integral notation (\ref{propagator1}), one can write
\begin{align}
&\frac{\partial}{\partial \lambda} K_N[ \lambda x,  \lambda y, \boldsymbol{0} ]  (0 , \beta | 0, 0 )\nonumber \\ & = - \int_{0,0}^{0, \beta } \mathcal{D} \overline{\mathbf{r}} \left( \frac{m}{2} \sum_{i}^N \int_0^{ \beta} d\tau \int_0^{\beta} d\sigma x(\tau - \sigma)  \mathbf{r_i}(\tau) \cdot \mathbf{r_i}(\sigma)  \right.  \nonumber  \\
  & \left.  + \frac{m}{2N}  \sum_{i,j} \int_0^{ \beta} d\tau \int_0^{\beta} d\sigma \left[y(\tau-\sigma)-x(\tau-\sigma) \right] \mathbf{r_i}(\tau) \cdot \mathbf{r_j}(\sigma) \right)   e^{-S^{(N)}[\overline{\mathbf{r}}, \lambda x , \lambda  y , \mathbf{0}]}  .
\end{align}

By making use of functional derivatives with respect to the source terms $\boldsymbol{\kappa}_i$ and taking them out of the path-integral, the propagator fraction in the $\lambda$-integral of (\ref{J1}) can be written as

\begin{align}
& \frac{\frac{\partial}{\partial \lambda}  K_N[ \lambda x,  \lambda y, \boldsymbol{0} ]  (0 , \beta | 0, 0 )}
{ K_N[ \lambda x,  \lambda y, \boldsymbol{0} ]  (0 , \beta | 0, 0 )}=-\left( \frac{1}{2m} \sum_{i}^N \int_0^{ \beta} d\tau \int_0^{\beta} d\sigma x(\tau - \sigma)   \frac{\delta}{\delta \boldsymbol{\kappa}_i(\tau)} \cdot \frac{\delta}{\delta \boldsymbol{\kappa}_i(\sigma)} \right. \nonumber \\
  & \left.  \left.  + \frac{1}{2Nm}   \int_0^{ \beta} d\tau \int_0^{\beta} d\sigma \left[y(\tau-\sigma)-x(\tau-\sigma) \right] \sum_i \frac{\delta}{\delta \boldsymbol{\kappa}_i(\tau)}   \cdot \sum_{j}  \frac{\delta}{\delta \boldsymbol{\kappa}_j(\sigma)}  \right) e^{ -S_{\textrm{cl}} \left[ \lambda x, \lambda y, \overline{\boldsymbol{\kappa} }\right] \left( 0, 0 \right)}  \right|_{  \boldsymbol{\overline{\kappa}} = 0   }   .
  \label{lambdapropagatorfaction}
\end{align}
Since $S_{\textrm{cl}} \left[ \lambda x, \lambda y, \overline{\boldsymbol{\kappa} }\right] \left( 0, 0 \right)$ is known, the functional derivatives can be straightforwardly performed to obtain:

\begin{align}
&  \frac{\frac{\partial}{\partial \lambda} K_N[\lambda x, \lambda y, \boldsymbol{0} ]  (0 , \beta | 0, 0 )}{K_N[\lambda x, \lambda y, \boldsymbol{0} ]  (0 , \beta | 0, 0 )}=   \frac{d}{2}	(N-1) \left[   \left(\sum_n \frac{1}{\nu_n^2 + \lambda \beta x_n} \right)^{-1} \sum_n \frac{ \beta x_n}{(\nu_n^2 + \lambda \beta x_n)^2} -  \sum_n \frac{\beta x_n}{\nu_n^2 + \lambda \beta x_n}  \right]   \nonumber \\
& + \frac{d}{2} \left[ \left(\sum_n \frac{1}{(\nu_n^2 + \lambda \beta y_n)^2} \right)^{-1} \sum_n \frac{\beta y_n}{(\nu_n^2 + \lambda \beta y_n)^2} -  \sum_n \frac{\beta y_n}{\nu_n^2 + \lambda \beta y_n} \right]  .
  \label{lambdapropagatorfaction2}
\end{align}

The $\lambda$-integral in (\ref{J1}) can now be analytically computed to finally obtain the many-body fluctuation factor:

\begin{equation}
K_N[x,y, \boldsymbol{0} ] (0 , \beta | 0, 0 ) = K[x, \boldsymbol{0} ]  (0 , \beta | 0, 0 )^{(N-1)} K[y, \boldsymbol{0} ]  (0 , \beta | 0, 0 ),
\label{fluctuationfactorization}
\end{equation}
where the single-particle fluctuation factor in $d$ dimensions is given by

\begin{equation}
K[x, \boldsymbol{0} ]  (0 , \beta | 0, 0 ) =  \left( \frac{m}{2 \pi \beta }\right) ^{\frac{d}{2} }  \left( \frac{4}{\beta^3 x_0 \Delta_x } \right)^{ \frac{d}{2}} \left( \frac{1}{\prod_{k=1}^{\infty} \left( 1 + \frac{\beta x_k}{\nu_k^2} \right) } \right)^{d}.
\end{equation}
This result together with (\ref{propagator2}) and (\ref{classicalaction4}) proves the factorization of the propagator in (\ref{propagatorfactorization1}).

\section{Explicit evaluation of closed loop Gaussian integrals}\label{Appendix B}

Let us start by defining a shorthand notation for the single-particle propagator (\ref{SingleParticlePropagator}) with $\boldsymbol{\kappa}_i=0$:

\begin{align}
K[x, \boldsymbol{0} ] (\mathbf{r}_T, \beta | \mathbf{r}_0, 0 ) &=   \mathcal{A}^d \exp \left( - a \left( \mathbf{r}_T - \mathbf{r}_0 \right)^2 - b \left( \mathbf{r}_T + \mathbf{r}_0 \right)^2 \right),
\label{SingleParticlePropagator2}
\end{align}
where $a= \frac{m}{2\beta} A_x$, $b=\frac{m}{2\beta} \frac{1}{\Delta_x}$, and

\begin{align}
\mathcal{A}&= \left( \frac{m}{2 \pi \beta} \right)^{1/2} \left(  \frac{4}{\beta^3 x_0 \Delta_x} \right)^{1/2} \frac{1}{ {\displaystyle \prod_{k=1}  \left(  1 + \frac{\beta x_k}{\nu_k^2}\right) } }.
\end{align}
It follows from expression (\ref{hq1}) and (\ref{SingleParticlePropagator2}) that the cyclic integral $h_\ell(\mathbf{k})$ factorizes as a product of each dimensional component $h_\ell(\mathbf{k})= h_\ell(k_x) h_\ell(k_y) h_\ell(k_z)$, where each factor is of the form:

\begin{equation}
h_\ell(k_z)= \mathcal{A}^{\ell} \int_{-\infty}^{\infty} dz_1  ... \int_{-\infty}^{\infty} dz_\ell K[x,0 ] ( z_1, \beta | z_\ell, 0 ) ... K[x,0 ] ( z_3, \beta | z_2, 0 )  K[x ,0] ( z_2, \beta | z_1 0 ) e^{ - i\frac{1}{N}   k_z \sum_{j=1}^{\ell} z_j}.
\label{hq2}
\end{equation}
Here, the notation for $K[x,0 ] ( z_1, \beta | z_\ell, 0 )$ as a function of scalar points $z_T$ and $z_0$ rather than vector variables refers to the propagator (\ref{SingleParticlePropagator2}) in one dimension $d=1$. After substitution of the propagators, expression (\ref{hq2}) can also be calculated using the well-known Gaussian integral formula:

\begin{equation}
h_\ell(k_z)= \mathcal{A}^{\ell} \int_{-\infty}^{\infty} dz_1  ... \int_{-\infty}^{\infty} dz_\ell  \exp \left( -\mathbf{z}^T \mathcal{C} \mathbf{z} - \mathbf{B}^T \mathbf{z} \right)= \mathcal{A}^\ell \sqrt{\frac{\pi^\ell}{\det ( \mathcal C )}} \exp \left( \frac{1}{4} \mathbf{B}^T \mathcal{C}^{-1} \mathbf{B} \right),
\label{hq3}
\end{equation}
where we invoke a vector notation for $\mathbf{z}^T = \left( z_1 , ... , z_\ell \right) $, $\mathbf{B}^T= \frac{ik_z}{N} \left(1,...,1 \right)$ and define the $\ell \times \ell$-dimensional matrix as

\begin{equation}
\mathcal{C}= \begin{pmatrix}
2(a+b) & (b-a) & 0 & ... & (b-a) \\
(b-a) & 2(a+b) & (b-a)   & ...& ... \\
0     &  (b-a) & 2(a+b) & ... & 0\\
...   & ...    & ... & ...  & (b-a) \\
(b-a) & ... & 0 & (b-a) & 2(a+b) 
\end{pmatrix} . 
\end{equation}

The matrix $\mathcal{C}$ is a circulant matrix, characterized by the property that any row or column is obtained by shifting the previous one by a single space (using periodic boundary conditions at the edges). 
Every circulant matrix has the same set of $j=\lbrace 0,1,...,\ell-1 \rbrace$ eigenvectors \cite{CirculantGray}:

\begin{equation}
\mathbf{y}_j^T= \frac{1}{\sqrt{\ell}} \left( \rho_j^0 , \rho_j^1, ..., \rho_j^{\ell-1} \right),  \hspace{10pt} \text{where} \hspace{5pt} \rho_j = e^{\frac{2 \pi i }{\ell} j} ,
\label{eigenvectors1}
\end{equation}
with corresponding eigenvalues for this particular matrix \cite{CirculantGray}:

\begin{equation}
\lambda_j = 2 (a+b) + 2(b-a) \cos( \frac{2 \pi j}{\ell} ).
\label{eigenvalues1}
\end{equation}
The goal now is to calculate both the determinant of $\mathcal{C}$ and the quadratic form $\mathbf{B}^T \mathcal{C}^{-1} \mathbf{B}$ of its inverse to obtain an explicit expression of (\ref{hq3}). 
An expression for the determinant is readily written down as the product over all eigenvalues:

\begin{equation}
\det(\mathcal{C}) = \prod_{j=0}^{\ell-1} \left(2 (a+b) + 2(b-a) \cos( \frac{2 \pi j}{\ell} ) \right)= \left[ 2 (a-b) \right] ^\ell  \prod_{j=0}^{\ell-1} \left( \frac{ a+b}{a-b} - \cos( \frac{2 \pi j}{\ell} ) \right) .
\label{determinantproduct1}
\end{equation}
Consider the strictly positive real numbers $a$ and $b$ and assume  $a \neq b$. 
We can now define $\tilde{z} = \textrm{arccosh}\left(\frac{a+b}{a-b} \right)$. For $\frac{a+b}{a-b}>1$, $\tilde{z}$ is real and uniquely defined. 
However, any $\frac{a+b}{a-b}<1$ lies exactly on the branch cut of the $\textrm{arccosh}$-function, and $\tilde{z}$ is complex and uniquely defined only up to the choice of whether the branch cut is approached from above or below the real axis. 
Either of the two choices work, and as we will show both yield the same result. Having converted $\frac{a+b}{a-b}$ in this form, the cosines in (\ref{determinantproduct1}) can now be added:

\begin{equation}
\det(\mathcal{C}) =  \left[ 2 (a-b) \right] ^\ell  \prod_{j=0}^{\ell-1} \left( \cos(i \tilde{z} ) - \cos( \frac{2 \pi j}{\ell} ) \right) =  \left[ 4 (a-b) \right] ^\ell  \prod_{j=0}^{\ell-1}  \sin(\frac{\pi j}{\ell} + \frac{i \tilde{z}}{2} ) \prod_{j=0}^{\ell-1} \sin(\frac{\pi j}{\ell} -\frac{i \tilde{z}}{2} )  .
\label{determinantC}
\end{equation}
We encountered a very concise proof of the resulting sine product series in \cite{MathSE}. First note that the following polynomial in $c$ can be decomposed in terms of its roots:

\begin{equation}
c^\ell - 1 =\prod_{j=0}^{\ell-1} \left(c-e^{\frac{2 \pi i}{\ell} j} \right) .
\end{equation}
Setting $c=e^{2iz}$, this can be applied to factorize the sine function as follows:

\begin{align}
\sin(\ell z) &=  \frac{e^{-i\ell z}}{2i} \left( e^{2i\ell z} - 1 \right) = \frac{e^{-i\ell z}}{2i} \prod_{j=0}^{\ell-1} \left(e^{2i z}-e^{\frac{2 \pi i}{\ell} j} \right). \label{sinqz}
\end{align}
After some algebraic manipulations on (\ref{sinqz}) one readily obtains for any complex $z$:

\begin{equation}
  \prod_{j=0}^{\ell-1}   \sin(\frac{\pi j }{\ell}  + z )= \frac{1}{2^{\ell-1}} \sin(\ell z),
\end{equation}
which is the known result found in tables of product series \cite{jeffrey2000table}. Using this result in (\ref{determinantC}) yields:

\begin{equation}
\det( \mathcal{C} ) =  4  (a-b)  ^\ell  \sinh( \frac{\ell}{2} \tilde{z} )^2. 
\label{determinant1}
\end{equation}
Let us now go back to the ambiguity of defining $\tilde{z}$ along the branch cut. If $-1<\frac{a+b}{a-b} <1$, then $\tilde{z}$ is purely imaginary and only changes sign across the branch cut, which clearly does not affect (\ref{determinant1}). 
If $ \frac{a+b}{a-b} < -1$, then the real part of $\tilde{z}$ remains constant along the branch cut and the imaginary part jumps from $\pi$ to $-\pi$, which does not change (\ref{determinant1}) for an integer $\ell$. 
Therefore any choice gives the same result, and we can unambiguously write:
\begin{equation}
\det( \mathcal{C} ) =  4  (a-b) ^\ell  \sinh( \frac{\ell}{2}  \textrm{arccosh}\left( \frac{a+b}{a-b} \right) )^2. 
\label{determinant2}
\end{equation}
We want to emphasize that when $a-b<0$, each of the two factors in (\ref{determinant2}) become negative for odd cycles $\ell$, but the determinant always remains strictly positive and hence the square root in (\ref{hq3}) is well defined and real.

Next we have to find the quadratic form of the inverse matrix $\mathbf{B}^T \mathcal{C}^{-1} \mathbf{B}$. 
For this, we note that the matrix $\mathcal{C}$ is diagonalized as $D= Q^* \mathcal{C} Q$ \cite{CirculantGray}, where $Q$ is the matrix with the normalized eigenvectors (\ref{eigenvectors1}), and $D$ is the matrix with eigenvalues (\ref{eigenvalues1}) on the diagonal. 
It readily follows that

\begin{equation}
\mathbf{B}^T \mathcal{C}^{-1} \mathbf{B} = \mathbf{B}^T Q D^{-1} Q^* \mathbf{B} = - \frac{k_z^2 \ell}{N^2} \frac{1}{4b}.
\label{quadraticform1}
\end{equation}
The determinant (\ref{determinant1}) and quadratic form of the inverse (\ref{quadraticform1}) now yield:

\begin{equation}
h_\ell(k_z) =  \mathcal{A}^\ell \left( \frac{\pi^\ell}{4  (a-b)   ^\ell  \sinh( \frac{\ell}{2}  \textrm{arccosh}\left( \frac{a+b}{a-b} \right) )^2} \right)^{1/2} \exp \left( - \frac{k_z^2 \ell}{N^2} \frac{1}{16b} \right).
\end{equation}
After substitution of $a$, $b$, $\mathcal{A}$, and taking the dimensionality into account, we exactly obtain expression (\ref{hqmain}) in section (\ref{Section 3}).


\begin{thebibliography}{23}%
\makeatletter
\providecommand \@ifxundefined [1]{%
 \@ifx{#1\undefined}
}%
\providecommand \@ifnum [1]{%
 \ifnum #1\expandafter \@firstoftwo
 \else \expandafter \@secondoftwo
 \fi
}%
\providecommand \@ifx [1]{%
 \ifx #1\expandafter \@firstoftwo
 \else \expandafter \@secondoftwo
 \fi
}%
\providecommand \natexlab [1]{#1}%
\providecommand \enquote  [1]{``#1''}%
\providecommand \bibnamefont  [1]{#1}%
\providecommand \bibfnamefont [1]{#1}%
\providecommand \citenamefont [1]{#1}%
\providecommand \href@noop [0]{\@secondoftwo}%
\providecommand \href [0]{\begingroup \@sanitize@url \@href}%
\providecommand \@href[1]{\@@startlink{#1}\@@href}%
\providecommand \@@href[1]{\endgroup#1\@@endlink}%
\providecommand \@sanitize@url [0]{\catcode `\\12\catcode `\$12\catcode
  `\&12\catcode `\#12\catcode `\^12\catcode `\_12\catcode `\%12\relax}%
\providecommand \@@startlink[1]{}%
\providecommand \@@endlink[0]{}%
\providecommand \url  [0]{\begingroup\@sanitize@url \@url }%
\providecommand \@url [1]{\endgroup\@href {#1}{\urlprefix }}%
\providecommand \urlprefix  [0]{URL }%
\providecommand \Eprint [0]{\href }%
\providecommand \doibase [0]{https://doi.org/}%
\providecommand \selectlanguage [0]{\@gobble}%
\providecommand \bibinfo  [0]{\@secondoftwo}%
\providecommand \bibfield  [0]{\@secondoftwo}%
\providecommand \translation [1]{[#1]}%
\providecommand \BibitemOpen [0]{}%
\providecommand \bibitemStop [0]{}%
\providecommand \bibitemNoStop [0]{.\EOS\space}%
\providecommand \EOS [0]{\spacefactor3000\relax}%
\providecommand \BibitemShut  [1]{\csname bibitem#1\endcsname}%
\let\auto@bib@innerbib\@empty
\bibitem [{\citenamefont {Feynman}(1955)}]{Feynman1955}%
  \BibitemOpen
  \bibfield  {author} {\bibinfo {author} {\bibfnamefont {R.~P.}\ \bibnamefont
  {Feynman}},\ }\bibfield  {title} {\bibinfo {title} {Slow electrons in a polar
  crystal},\ }\href {https://doi.org/10.1103/PhysRev.97.660} {\bibfield
  {journal} {\bibinfo  {journal} {Phys. Rev.}\ }\textbf {\bibinfo {volume}
  {97}},\ \bibinfo {pages} {660} (\bibinfo {year} {1955})}\BibitemShut
  {NoStop}%
\bibitem [{\citenamefont {Feynman}(1998)}]{feynman1998statistical}%
  \BibitemOpen
  \bibfield  {author} {\bibinfo {author} {\bibfnamefont {R.}~\bibnamefont
  {Feynman}},\ }\href@noop {} {\emph {\bibinfo {title} {Statistical Mechanics:
  A Set of Lectures}}},\ Advanced Books Classics\ (\bibinfo  {publisher}
  {Avalon, New York},\ \bibinfo {year} {1998})\BibitemShut {NoStop}%
\bibitem [{\citenamefont {Feynman}\ and\ \citenamefont
  {Vernon}(1963)}]{Feynman1963}%
  \BibitemOpen
  \bibfield  {author} {\bibinfo {author} {\bibfnamefont {R.~P.}\ \bibnamefont
  {Feynman}}\ and\ \bibinfo {author} {\bibfnamefont {F.~L.}\ \bibnamefont
  {Vernon}},\ }\bibfield  {title} {\bibinfo {title} {The theory of a general
  quantum system interacting with a linear dissipative system},\ }\href
  {https://doi.org/10.1016/0003-4916(63)90068-x} {\bibfield  {journal}
  {\bibinfo  {journal} {Annals of Physics}\ }\textbf {\bibinfo {volume} {24}},\
  \bibinfo {pages} {1208} (\bibinfo {year} {1963})}\BibitemShut {NoStop}%
\bibitem [{\citenamefont {Caldeira}\ and\ \citenamefont
  {Leggett}(1983)}]{Caldeira1983}%
  \BibitemOpen
  \bibfield  {author} {\bibinfo {author} {\bibfnamefont {A.}~\bibnamefont
  {Caldeira}}\ and\ \bibinfo {author} {\bibfnamefont {A.}~\bibnamefont
  {Leggett}},\ }\bibfield  {title} {\bibinfo {title} {Path integral approach to
  quantum Brownian motion},\ }\href
  {https://doi.org/https://doi.org/10.1016/0378-4371(83)90013-4} {\bibfield
  {journal} {\bibinfo  {journal} {Physica A: Statistical Mechanics and its
  Applications}\ }\textbf {\bibinfo {volume} {121}},\ \bibinfo {pages} {587}
  (\bibinfo {year} {1983})}\BibitemShut {NoStop}%
\bibitem [{\citenamefont {Rosenfelder}\ and\ \citenamefont
  {Schreiber}(2001)}]{Rosenfelder2001}%
  \BibitemOpen
  \bibfield  {author} {\bibinfo {author} {\bibfnamefont {R.}~\bibnamefont
  {Rosenfelder}}\ and\ \bibinfo {author} {\bibfnamefont {A.}~\bibnamefont
  {Schreiber}},\ }\bibfield  {title} {\bibinfo {title} {On the best quadratic
  approximation in Feynman's path integral treatment of the polaron},\ }\href
  {https://doi.org/https://doi.org/10.1016/S0375-9601(01)00287-0} {\bibfield
  {journal} {\bibinfo  {journal} {Physics Letters A}\ }\textbf {\bibinfo
  {volume} {284}},\ \bibinfo {pages} {63} (\bibinfo {year} {2001})}\BibitemShut
  {NoStop}%
\bibitem [{\citenamefont {Verbist}\ \emph {et~al.}(1991)\citenamefont
  {Verbist}, \citenamefont {Peeters},\ and\ \citenamefont
  {Devreese}}]{Verbist1992}%
  \BibitemOpen
  \bibfield  {author} {\bibinfo {author} {\bibfnamefont {G.}~\bibnamefont
  {Verbist}}, \bibinfo {author} {\bibfnamefont {F.~M.}\ \bibnamefont
  {Peeters}},\ and\ \bibinfo {author} {\bibfnamefont {J.~T.}\ \bibnamefont
  {Devreese}},\ }\bibfield  {title} {\bibinfo {title} {Large bipolarons in two
  and three dimensions},\ }\href {https://doi.org/10.1103/PhysRevB.43.2712}
  {\bibfield  {journal} {\bibinfo  {journal} {Phys. Rev. B}\ }\textbf {\bibinfo
  {volume} {43}},\ \bibinfo {pages} {2712} (\bibinfo {year}
  {1991})}\BibitemShut {NoStop}%
\bibitem [{\citenamefont {Casteels}\ \emph {et~al.}(2013)\citenamefont
  {Casteels}, \citenamefont {Tempere},\ and\ \citenamefont
  {Devreese}}]{Casteels2013}%
  \BibitemOpen
  \bibfield  {author} {\bibinfo {author} {\bibfnamefont {W.}~\bibnamefont
  {Casteels}}, \bibinfo {author} {\bibfnamefont {J.}~\bibnamefont {Tempere}},\
  and\ \bibinfo {author} {\bibfnamefont {J.~T.}\ \bibnamefont {Devreese}},\
  }\bibfield  {title} {\bibinfo {title} {Bipolarons and multipolarons
  consisting of impurity atoms in a Bose-Einstein condensate},\ }\href
  {https://doi.org/10.1103/PhysRevA.88.013613} {\bibfield  {journal} {\bibinfo
  {journal} {Phys. Rev. A}\ }\textbf {\bibinfo {volume} {88}},\ \bibinfo
  {pages} {013613} (\bibinfo {year} {2013})}\BibitemShut {NoStop}%
\bibitem [{\citenamefont {Klimin}\ \emph {et~al.}(2004)\citenamefont {Klimin},
  \citenamefont {Fomin}, \citenamefont {Brosens},\ and\ \citenamefont
  {Devreese}}]{Klimin2004}%
  \BibitemOpen
  \bibfield  {author} {\bibinfo {author} {\bibfnamefont {S.~N.}\ \bibnamefont
  {Klimin}}, \bibinfo {author} {\bibfnamefont {V.~M.}\ \bibnamefont {Fomin}},
  \bibinfo {author} {\bibfnamefont {F.}~\bibnamefont {Brosens}},\ and\ \bibinfo
  {author} {\bibfnamefont {J.~T.}\ \bibnamefont {Devreese}},\ }\bibfield
  {title} {\bibinfo {title} {Ground state and optical conductivity of
  interacting polarons in a quantum dot},\ }\href
  {https://doi.org/10.1103/PhysRevB.69.235324} {\bibfield  {journal} {\bibinfo
  {journal} {Phys. Rev. B}\ }\textbf {\bibinfo {volume} {69}},\ \bibinfo
  {pages} {235324} (\bibinfo {year} {2004})}\BibitemShut {NoStop}%
\bibitem [{\citenamefont {Brosens}\ \emph
  {et~al.}(1997{\natexlab{a}})\citenamefont {Brosens}, \citenamefont
  {Devreese},\ and\ \citenamefont {Lemmens}}]{Brosens1997a}%
  \BibitemOpen
  \bibfield  {author} {\bibinfo {author} {\bibfnamefont {F.}~\bibnamefont
  {Brosens}}, \bibinfo {author} {\bibfnamefont {J.~T.}\ \bibnamefont
  {Devreese}},\ and\ \bibinfo {author} {\bibfnamefont {L.~F.}\ \bibnamefont
  {Lemmens}},\ }\bibfield  {title} {\bibinfo {title} {Thermodynamics of coupled
  identical oscillators within the path-integral formalism},\ }\href
  {https://doi.org/10.1103/PhysRevE.55.227} {\bibfield  {journal} {\bibinfo
  {journal} {Phys. Rev. E}\ }\textbf {\bibinfo {volume} {55}},\ \bibinfo
  {pages} {227} (\bibinfo {year} {1997}{\natexlab{a}})}\BibitemShut {NoStop}%
\bibitem [{\citenamefont {Brosens}\ \emph
  {et~al.}(1997{\natexlab{b}})\citenamefont {Brosens}, \citenamefont
  {Devreese},\ and\ \citenamefont {Lemmens}}]{Brosens1997b}%
  \BibitemOpen
  \bibfield  {author} {\bibinfo {author} {\bibfnamefont {F.}~\bibnamefont
  {Brosens}}, \bibinfo {author} {\bibfnamefont {J.~T.}\ \bibnamefont
  {Devreese}},\ and\ \bibinfo {author} {\bibfnamefont {L.~F.}\ \bibnamefont
  {Lemmens}},\ }\bibfield  {title} {\bibinfo {title} {Density and pair
  correlation function of confined identical particles: The Bose-Einstein
  case},\ }\href {https://doi.org/10.1103/PhysRevE.55.6795} {\bibfield
  {journal} {\bibinfo  {journal} {Phys. Rev. E}\ }\textbf {\bibinfo {volume}
  {55}},\ \bibinfo {pages} {6795} (\bibinfo {year}
  {1997}{\natexlab{b}})}\BibitemShut {NoStop}%
\bibitem [{\citenamefont {Ingold}\ \emph {et~al.}(2009)\citenamefont {Ingold},
  \citenamefont {H\"anggi},\ and\ \citenamefont {Talkner}}]{Ingold2009}%
  \BibitemOpen
  \bibfield  {author} {\bibinfo {author} {\bibfnamefont {G.-L.}\ \bibnamefont
  {Ingold}}, \bibinfo {author} {\bibfnamefont {P.}~\bibnamefont {H\"anggi}},\
  and\ \bibinfo {author} {\bibfnamefont {P.}~\bibnamefont {Talkner}},\
  }\bibfield  {title} {\bibinfo {title} {Specific heat anomalies of open
  quantum systems},\ }\href {https://doi.org/10.1103/PhysRevE.79.061105}
  {\bibfield  {journal} {\bibinfo  {journal} {Phys. Rev. E}\ }\textbf {\bibinfo
  {volume} {79}},\ \bibinfo {pages} {061105} (\bibinfo {year}
  {2009})}\BibitemShut {NoStop}%
\bibitem [{\citenamefont {Adamietz}\ \emph {et~al.}(2014)\citenamefont
  {Adamietz}, \citenamefont {Ingold},\ and\ \citenamefont
  {Weiss}}]{Ingold2014}%
  \BibitemOpen
  \bibfield  {author} {\bibinfo {author} {\bibfnamefont {R.}~\bibnamefont
  {Adamietz}}, \bibinfo {author} {\bibfnamefont {G.-L.}\ \bibnamefont
  {Ingold}},\ and\ \bibinfo {author} {\bibfnamefont {U.}~\bibnamefont
  {Weiss}},\ }\bibfield  {title} {\bibinfo {title} {Thermodynamic anomalies in
  the presence of general linear dissipation: from the free particle to the
  harmonic oscillator},\ }\href@noop {} {\bibfield  {journal} {\bibinfo
  {journal} {The European Physical Journal B}\ }\textbf {\bibinfo {volume}
  {87}},\ \bibinfo {pages} {90} (\bibinfo {year} {2014})}\BibitemShut {NoStop}%
\bibitem [{\citenamefont {Mahan}(2000)}]{mahan}%
  \BibitemOpen
  \bibfield  {author} {\bibinfo {author} {\bibfnamefont {G.}~\bibnamefont
  {Mahan}},\ }\href {https://books.google.be/books?id=xzSgZ4-yyMEC} {\emph
  {\bibinfo {title} {Many-Particle Physics}}},\ Physics of Solids and Liquids\
  (\bibinfo  {publisher} {Springer, New York},\ \bibinfo {year} {2000})\BibitemShut
  {NoStop}%
\bibitem [{\citenamefont {Tempere}\ \emph {et~al.}(2009)\citenamefont
  {Tempere}, \citenamefont {Casteels}, \citenamefont {Oberthaler},
  \citenamefont {Knoop}, \citenamefont {Timmermans},\ and\ \citenamefont
  {Devreese}}]{Tempere2009}%
  \BibitemOpen
  \bibfield  {author} {\bibinfo {author} {\bibfnamefont {J.}~\bibnamefont
  {Tempere}}, \bibinfo {author} {\bibfnamefont {W.}~\bibnamefont {Casteels}},
  \bibinfo {author} {\bibfnamefont {M.~K.}\ \bibnamefont {Oberthaler}},
  \bibinfo {author} {\bibfnamefont {S.}~\bibnamefont {Knoop}}, \bibinfo
  {author} {\bibfnamefont {E.}~\bibnamefont {Timmermans}},\ and\ \bibinfo
  {author} {\bibfnamefont {J.~T.}\ \bibnamefont {Devreese}},\ }\bibfield
  {title} {\bibinfo {title} {Feynman path-integral treatment of the
  BEC-impurity polaron},\ }\href {https://doi.org/10.1103/PhysRevB.80.184504}
  {\bibfield  {journal} {\bibinfo  {journal} {Phys. Rev. B}\ }\textbf {\bibinfo
  {volume} {80}},\ \bibinfo {pages} {184504} (\bibinfo {year}
  {2009})}\BibitemShut {NoStop}%
\bibitem [{\citenamefont {Ichmoukhamedov}\ and\ \citenamefont
  {Tempere}(2019)}]{ExtendedFrohlichFeynman2019}%
  \BibitemOpen
  \bibfield  {author} {\bibinfo {author} {\bibfnamefont {T.}~\bibnamefont
  {Ichmoukhamedov}}\ and\ \bibinfo {author} {\bibfnamefont {J.}~\bibnamefont
  {Tempere}},\ }\bibfield  {title} {\bibinfo {title} {Feynman path-integral
  treatment of the Bose polaron beyond the Fr\"ohlich model},\ }\href
  {https://doi.org/10.1103/PhysRevA.100.043605} {\bibfield  {journal} {\bibinfo
   {journal} {Phys. Rev. A}\ }\textbf {\bibinfo {volume} {100}},\ \bibinfo
  {pages} {043605} (\bibinfo {year} {2019})}\BibitemShut {NoStop}%
\bibitem [{\citenamefont {Houtput}\ and\ \citenamefont
  {Tempere}(2020)}]{houtput2020beyondfrohlich}%
  \BibitemOpen
  \bibfield  {author} {\bibinfo {author} {\bibfnamefont {M.}~\bibnamefont
  {Houtput}}\ and\ \bibinfo {author} {\bibfnamefont {J.}~\bibnamefont
  {Tempere}},\ }\href@noop {} {\bibinfo {title} {Beyond the Fr\"ohlich Hamiltonian: Path-integral treatment of large polarons in anharmonic solids}}
  \href
  {https://doi.org/10.1103/PhysRevB.103.184306} {\bibfield  {journal} {\bibinfo
   {journal} {Phys. Rev. B}\ }\textbf {\bibinfo {volume} {103}},\ \bibinfo
  {pages} {184306} (\bibinfo {year} {2021})}\BibitemShut {NoStop}%
\bibitem [{\citenamefont {Adamowski}\ \emph {et~al.}(1982)\citenamefont
  {Adamowski}, \citenamefont {Gerlach},\ and\ \citenamefont
  {Leschke}}]{Adamowski}%
  \BibitemOpen
  \bibfield  {author} {\bibinfo {author} {\bibfnamefont {J.}~\bibnamefont
  {Adamowski}}, \bibinfo {author} {\bibfnamefont {B.}~\bibnamefont {Gerlach}},\
  and\ \bibinfo {author} {\bibfnamefont {H.}~\bibnamefont {Leschke}},\
  }\bibfield  {title} {\bibinfo {title} {Explicit evaluation of certain
  Gaussian functional integrals arising in problems of statistical physics},\
  }\href {https://doi.org/10.1063/1.525343} {\bibfield  {journal} {\bibinfo
  {journal} {Journal of Mathematical Physics}\ }\textbf {\bibinfo {volume}
  {23}},\ \bibinfo {pages} {243} (\bibinfo {year} {1982})}\   \BibitemShut {NoStop}%
\bibitem [{\citenamefont {Hasegawa}(2011{\natexlab{a}})}]{Hasegawa1}%
  \BibitemOpen
  \bibfield  {author} {\bibinfo {author} {\bibfnamefont {H.}~\bibnamefont
  {Hasegawa}},\ }\bibfield  {title} {\bibinfo {title} {Classical small systems
  coupled to finite baths},\ }\href
  {https://doi.org/10.1103/PhysRevE.83.021104} {\bibfield  {journal} {\bibinfo
  {journal} {Phys. Rev. E}\ }\textbf {\bibinfo {volume} {83}},\ \bibinfo
  {pages} {021104} (\bibinfo {year} {2011}{\natexlab{a}})}\BibitemShut
  {NoStop}%
\bibitem [{\citenamefont {Hasegawa}(2011{\natexlab{b}})}]{Hasegawa2}%
  \BibitemOpen
  \bibfield  {author} {\bibinfo {author} {\bibfnamefont {H.}~\bibnamefont
  {Hasegawa}},\ }\bibfield  {title} {\bibinfo {title} {Specific heat anomalies
  of small quantum systems subjected to finite baths},\ }\href
  {https://doi.org/10.1063/1.3669485} {\bibfield  {journal} {\bibinfo
  {journal} {Journal of Mathematical Physics}\ }\textbf {\bibinfo {volume}
  {52}},\ \bibinfo {pages} {123301} (\bibinfo {year} {2011}{\natexlab{b}})}\  \BibitemShut {NoStop}%
\bibitem [{\citenamefont {Ingold}(2012)}]{Ingold2012}%
  \BibitemOpen
  \bibfield  {author} {\bibinfo {author} {\bibfnamefont {G.-L.}\ \bibnamefont
  {Ingold}},\ }\bibfield  {title} {\bibinfo {title} {Thermodynamic anomaly of
  the free damped quantum particle: the bath perspective},\ }\href
  {https://doi.org/10.1140/epjb/e2011-20930-2} {\bibfield  {journal} {\bibinfo
  {journal} {The European Physical Journal B}\ }\textbf {\bibinfo {volume}
  {85}},\ \bibinfo {pages} {30} (\bibinfo {year} {2012})}\BibitemShut {NoStop}%
\bibitem [{\citenamefont {Gray}(2006)}]{CirculantGray}%
  \BibitemOpen
  \bibfield  {author} {\bibinfo {author} {\bibfnamefont {R.~M.}\ \bibnamefont
  {Gray}},\ }\bibfield  {title} {\bibinfo {title} {Toeplitz and circulant
  matrices: A review},\ }\href {https://doi.org/10.1561/0100000006} {\bibfield
  {journal} {\bibinfo  {journal} {Foundations and Trends in Communications and
  Information Theory}\ }\textbf {\bibinfo {volume} {2}},\ \bibinfo {pages}
  {155} (\bibinfo {year} {2006})}\BibitemShut {NoStop}%
\bibitem [{\citenamefont {(https://math.stackexchange.com/users/218419/mark
  viola)}()}]{MathSE}%
  \BibitemOpen
  \bibfield  {author} {\bibinfo {author} {\bibfnamefont {M. Viola.}\  },\ }\href
  {https://math.stackexchange.com/q/2096321} {\bibinfo {title} {Product
  identity multiple angle or
  $\sin(nx)=2^{n-1}\prod_{k=0}^{n-1}\sin\left(\frac{k\pi}n+x\right)$}},\
  \bibinfo {howpublished} {Mathematics Stack Exchange},\ \bibinfo {note}
  {(Dec. 10, 2017)},\
  {https://math.stackexchange.com/q/2096321} \BibitemShut {NoStop}%
\bibitem [{\citenamefont {Jeffrey}\ and\ \citenamefont
  {Zwillinger}(2000)}]{jeffrey2000table}%
  \BibitemOpen
  \bibfield  {author} {\bibinfo {author} {\bibfnamefont {A.}~\bibnamefont
  {Jeffrey}}\ and\ \bibinfo {author} {\bibfnamefont {D.}~\bibnamefont
  {Zwillinger}},\ }\href {https://books.google.gm/books?id=h4y-36vKIZgC} {\emph
  {\bibinfo {title} {Table of Integrals, Series, and Products}}}\ (\bibinfo
  {publisher} {Elsevier Science, Amsterdam},\ \bibinfo {year} {2000})\BibitemShut
  {NoStop}%
\end{thebibliography}
\end{document}